\documentclass[11pt]{article}
\usepackage{amsmath,amssymb,mathtools,bm}
\usepackage{microtype,booktabs,array,xcolor,graphicx}
\usepackage[numbers,sort&compress]{natbib}
\usepackage{hyperref,cleveref}
\graphicspath{{figures/}}

\title{Neural-network-powered new-physics searches\\
       by predictive goodness of fit}
\author{Jing Lei\\
Department of Statistics \& Data Science\\
Carnegie Mellon University\\
\texttt{jinglei@andrew.cmu.edu}}
\date{}

\begin{document}
\maketitle

\begin{abstract}
Model-independent searches for new physics test whether data agree with a
Reference model with unknown nuisance parameters. Because the
shape and location of a signal are not specified, these tests call for flexible
learners (fitting procedures such as neural networks), and a likelihood ratio maximized over a
neural network falls outside the standard asymptotic theory, making the calibration
a major challenge. We propose a predictive alternative: the departure from the
Reference is fitted on part of the data and its likelihood gain over the fitted
Reference is evaluated on the rest, cross-fitted over several folds to preserve
sample efficiency. The nuisance estimation error is
propagated into the variance of the test statistic to first order,
in the spirit of Neyman's $C(\alpha)$ test,
so that the test need not be
recalibrated at each nuisance value. The randomness in the fitted departure
is treated by a
central limit theorem for cross-validated statistics under algorithmic
stability, plus a variance correction when stability is not achieved.
The calibration is analytic, and no pseudo-experiments are needed.
On
established one-dimensional and five-dimensional (dilepton) benchmarks, the
false-positive rate always stays within Monte Carlo error of nominal across different
nuisances and the learners' tuning. The test's power relative to existing learned tests
is mixed: higher for some signals and lower for others, depending on
how well the fitted departure captures the signal.
\end{abstract}

\section{Introduction}\label{sec:intro}

Modern searches for physics beyond the Standard Model (SM) ask a broad question: is the
observed data set compatible with the SM prediction at all? Statistically this
is a goodness-of-fit test for a Reference model known only up to nuisance parameters that are
constrained by noisy auxiliary measurements.
A useful test for this task must meet two requirements: (i) it
must be sensitive to departures whose location and shape are not specified in
advance, which in practice means fitting a flexible model of the departure; and (ii)
its false-positive rate must be controlled when the nuisance parameters are
estimated.

The New Physics Learning Machine (NPLM)~\cite{DAgnolo2019learning,DAgnolo2021learning,DAgnolo2022learning}
meets the first requirement by fitting the departure with a neural network whose
loss is the negative logarithm of the likelihood ratio, and the test statistic
$t$ is twice the maximized log-likelihood ratio, with the nuisances profiled
against their auxiliary likelihoods. The second requirement is met through
Wilks' theorem: $t$ is expected to follow a $\chi^2$ distribution whose number
of degrees of freedom equals the number of network parameters. Our comparisons
are with its implementation with nuisances, Ref.~\cite{DAgnolo2022learning}.

This calibration has two weaknesses. The first is that the conditions of
Wilks' theorem are unlikely to hold for the neural network fit. The
theorem describes the maximized likelihood ratio of a regular parametric
family: the parameters must be identifiable, the true value must lie in the
interior of the parameter space, the likelihood must be well approximated by a
quadratic around its maximum, and the number of parameters must be small
compared with the information in the data. A neural network violates each of
these: for example, at the null hypothesis its hidden-layer parameters have no
effect on the likelihood, the situation in which the likelihood ratio is known
not to follow Wilks' $\chi^2$~\cite{Davies1987hypothesis,Gross2010trial}, and
weight clipping places parameters on the boundary of the allowed
region~\cite{Chernoff1954distribution,Self1987asymptotic}.
Departures from the $\chi^2$ law were noted from the
start: the original NPLM paper~\cite{DAgnolo2019learning} reports them in a
two-dimensional example, attributes one to ``insufficient training'', and
remarks that the agreement it does observe ``is not a consistency check''.
The $\chi^2$ law is enforced empirically: the weight-clipping bound is selected
on Reference-only pseudo-experiments (toys) so that the distribution of $t$
matches the target. This calibration is also fragile. In the multivariate benchmark of Ref.~\cite{DAgnolo2022learning},
clipping values that differ from the selected one at the second digit give
distributions of $t$ appreciably different from the target.
This leads to the second weakness: the regularization of the learner is also its calibration. The same
knob (e.g., weight clipping) decides how well the network can describe a signal and whether the
reported $p$-value is correct. This limits the freedom to choose a learner for the best power.

In this paper we address both issues by comparing predictions on held-out data.
The departure is still fitted with a flexible learner (any fitting procedure,
such as a penalized regression or a neural network), but it is fitted on one
part of the data and evaluated on the other, so that the statistic measures how
much better the fitted alternative predicts events it has not seen than the
fitted null does. Overfitting then lowers the statistic instead of inflating it,
and its null distribution no longer depends on the learner being in a $\chi^2$
regime. Such split-sample comparisons have been used
before~\cite{Wasserman2020universal,Chakravarti2023model}, but used as they are
they lose power in two ways. Our two main methodological contributions remove
these losses.
\begin{itemize}
\item \emph{Cross-fitting.} A single split spends events on the fit that are then
  not evaluated. Ref.~\cite{Grosso2024goodness} argued that ``the fit (i.e.,
  the training) must be performed with the very same data used to test'' and
  found that a 50/50 split costs NPLM as much power as discarding half the
  data. We cross-fit over $K$ folds, so that every event is used for both fitting and
  evaluation and none is discarded.
\item \emph{Adaptive null centering.} A held-out statistic compared with a fixed
  threshold, as in the split likelihood-ratio test of
  Ref.~\cite{Wasserman2020universal}, ignores that its null mean is negative for
  any fitted alternative, as Ref.~\cite{Grosso2024goodness} observed for NPLM
  evaluated on held-out data. We center it at its expectation under the fitted null,
  which adapts to the unknown null distribution, and studentize it.
\end{itemize}
The resulting statistic is calibrated analytically, with a normal limiting
distribution under the null hypothesis: its centering and variance are computed
from the fitted null, with explicit corrections for the estimation of the
nuisance parameters, for the finite Reference sample, and for the randomness of
the fits themselves.

The conditions behind this calibration are also weaker than those of Wilks'
theorem. The starting point is a central limit theorem for
cross-validated statistics~\cite{Lei2025modern,Austern2020asymptotics,Bayle2020cross},
whose main requirement on the
learner is stability: moving a single training event should change the fitted
departure by a small amount~\cite{Kearns1999algorithmic,Bousquet2002stability}.
This is a genuine condition on the learner, but one that is agnostic about how
the learner is parametrized and how its optimization is carried out.
Its parameters need not be
identifiable or lie in the interior of a parameter space, its fit need not reach
a maximum of the likelihood, and the number of parameters is not counted, so
clipped, penalized and early-stopped fits are covered as they are.
Regularization still matters for the calibration, but through stability, which
it improves, rather than by restricting the learner to a narrow regime. The
witness, our name for the fitted departure, can then be chosen for power,
and we use two very different ones: a penalized dictionary of smooth functions and Gaussian bumps, and the
same small neural networks as NPLM.
Whether a given learner is stable enough depends on the learner and on the
data. The precise conditions are stated in Section~\ref{sec:test-theory}.
In our experiments the stability condition did not hold under the null
hypothesis: with no departure to learn, the learner fits noise alone, so the
fitted witness varies from fit to fit by as much as its own size.
To address this, we add an explicit first-order variance correction for the randomness of
the fits, estimated from the fits themselves. Asymptotic theory for this regime
is yet to be developed, and the validity of the corrected test is demonstrated in
numerical experiments.

We test the method on the benchmarks of Ref.~\cite{DAgnolo2022learning}: a
one-dimensional exponential Reference with three signal types, and the
five-dimensional dilepton final state with a $Z'$ resonance and an
effective-field-theory (EFT) deformation, both with normalization and scale
nuisances. Two findings organize the results. First, the false-positive rate
stays close to its nominal value across true nuisance values one standard
deviation away from the central value, across the tuning of the learner (weight clipping,
training length, penalty strength), and across the two kinds of witness. Second, the power depends on how well the witness
captures the signal: in the dilepton benchmark our network is less sensitive than NPLM to
the narrow $Z'$ resonance and more sensitive to the broad EFT deformation.

The paper is organized as follows. Section~\ref{sec:setup} states the problem in
NPLM's notation. Section~\ref{sec:test} describes the test and its calibration,
and summarizes its guarantees and assumptions.
Section~\ref{sec:witness} describes the two witnesses.
Sections~\ref{sec:1d} and~\ref{sec:5d} present the benchmarks, and Section~\ref{sec:discussion} discusses limitations and
extensions. The appendices collect the calibration formulas, the statements of
the limit theorems with proof sketches, implementation details, and additional
results.

\section{Statistical setup}\label{sec:setup}

We use the notation of Ref.~\cite{DAgnolo2022learning} wherever possible;
Table~\ref{tab:notation} lists the correspondence.

\paragraph{Data.} The data set $\mathcal D=\{X_1,\dots,X_N\}$ consists of events
with features $x$ in a sample space $\mathcal X$, modeled as a Poisson process
with intensity $n(x)$: the number of events, $N$, is Poisson with mean $\int n$ and,
given $N$, the events are independent with density $n/\!\int n$. The Reference
(Standard Model) hypothesis predicts the intensity $n(x\,|\,\mathrm R_\nu)$ up to
a vector of nuisance parameters $\nu$, with $\nu=0$ the central value and
$N_0=\int n(\cdot\,|\,\mathrm R_0)$ the expected number of events there. The
construction applies to any finite set of nuisances; for concreteness we write
it for the two that are most common, $\nu=(\nu_n,\nu_s)$: $\nu_n$, a
log-normalization of the number of events produced, and $\nu_s$, a log-scale of
the measured features.

\paragraph{Auxiliary measurements.} The nuisances are constrained by an
independent auxiliary measurement $A$, summarized by an estimator
$\hat\nu_A\sim N(\nu^*,\Sigma_A)$ centered at the true value $\nu^*$, with
$\Sigma_A=\mathrm{diag}(\sigma_n^2,\sigma_s^2)$. Its log-likelihood is
\begin{equation}\label{eq:aux}
\ell_A(\nu)=-\tfrac12(\nu-\hat\nu_A)^\top\Sigma_A^{-1}(\nu-\hat\nu_A).
\end{equation}
The size of $\Sigma_A$ sets how much the data themselves must be used to
determine $\nu$: as $\Sigma_A\to0$ the Reference intensity is known exactly,
and as $\Sigma_A\to\infty$ the nuisances are determined by the data alone.

\paragraph{Access to the Reference model.} We distinguish two ways in which
the Reference model can be available, and the test is built to need only the
weaker one.
\begin{itemize}
\item \emph{Oracle access}: the intensity $n(x\,|\,\mathrm R_\nu)$ is known in
  analytical form for every $\nu$, and every expectation under the null can be
  computed by numerical integration.
\item \emph{Sample access}: the intensity is defined by a simulation chain and is
  available only through a Reference sample $\mathcal R$ of $M$ simulated events
  $X'_e$ at $\nu=0$, each carrying weight $r=N_0/M$, so that $\mathcal R$
  reproduces the expected count $N_0$.
  The scale nuisance acts on simulated events through a known transformation
  $T_{\nu_s}$ (for example a rescaling of measured momenta), and the normalization
  multiplies the weights by $e^{\nu_n}$, so
  samples at any $\nu$ are available.
\end{itemize}
Oracle access is the limit of sample access with an infinitely large Reference
sample, and the construction below is written so that the same
formulas cover both.

\paragraph{Hypotheses and validity.} The null hypothesis is that the data
follow the Reference model at some value of the nuisances,
\begin{equation}\label{eq:H0}
H_0:\quad n(\cdot)=n(\cdot\,|\,\mathrm R_{\nu^*})\quad\text{for some }\nu^*,
\end{equation}
where $\nu^*$ is also the value around which $\hat\nu_A$ is distributed. The
alternative is any other intensity; no form of the departure is assumed. A test
at nominal level $\alpha\in(0,1)$ is \emph{valid} if its rejection rate under $H_0$,
over repetitions of the data, the Reference sample and the auxiliary measurement,
is at most $\alpha$ for every value of $\nu^*$.

\begin{table}[t]
\centering
\caption{Notation used here and in Ref.~\cite{DAgnolo2022learning}.}
\label{tab:notation}
\small
\begin{tabular}{>{\raggedright\arraybackslash}p{0.34\textwidth}>{\raggedright\arraybackslash}p{0.3\textwidth}>{\raggedright\arraybackslash}p{0.26\textwidth}}
\toprule
quantity & this paper & Ref.~\cite{DAgnolo2022learning} \\
\midrule
data set, number of events & $\mathcal D$, $N$ & $\mathcal D$, $\mathcal N_{\mathcal D}$ \\
Reference intensity at nuisance $\nu$ & $n(x\,|\,\mathrm R_\nu)$ & $n(x\,|\,\mathrm R_\nu)$ \\
expected events at the central value & $N_0$ & $N(\mathrm R_0)$ \\
Reference sample, size, event weight & $\mathcal R$, $M$, $r=N_0/M$ & $\mathcal R$, $N_{\mathcal R}$, $w_e$ \\
nuisances (normalization, scale) & $\nu=(\nu_n,\nu_s)$ & $\nu=(\nu_{\rm N},\nu_{\rm S})$ \\
auxiliary data, estimate, uncertainty & $A$, $\hat\nu_A$, $\sigma_{n,s}$ & $\mathcal A$, $\hat\nu$, $\sigma_{\rm N,S}$ \\
nuisance log-ratio (learned approximation) & $h(x;\nu)$ ($\hat\delta_{1,2}$, Eq.~\eqref{eq:delta}) & $\log r(x;\nu)$ ($\log\hat r$) \\
fitted departure & witness $\phi_k$, fitted on all folds but $k$ & network $f(x;\mathbf w)$, fitted on $\mathcal D$ \\
test statistic & standardized $Z$ & $t=\tau-\Delta$ \\
\bottomrule
\end{tabular}
\end{table}

\section{The test}\label{sec:test}

\subsection{Cross-fitting and the fits on the training folds}\label{sec:test-idea}

The data events are assigned at random to $K$ folds
$\mathcal D_1,\dots,\mathcal D_K$ whose sizes differ by at most one, and with
sample access the
Reference sample is partitioned independently into folds
$\mathcal R_1,\dots,\mathcal R_K$. In our derivation we assume the fold sizes are equal for simplicity.
For a function $u$ of $x$ and a nuisance value $\nu$ we write
\[
\mathbb E_\nu[u]=\int n(\cdot\,|\,\mathrm R_\nu)\,u,\qquad
\mathbb E_{k,\nu}[u]=\frac1K\,\mathbb E_\nu[u],\qquad
\mathbb E^{\rm tr}_{k,\nu}[u]=\frac{K-1}K\,\mathbb E_\nu[u]
\]
for the expected sum of $u$ over all data events, over the held-out events of
fold $k$ and over its training events, when the nuisances take the value $\nu$.
With oracle access they are computed by numerical integration; with sample
access they are estimated by weighted sums over the Reference events
transformed by $T_{\nu_s}$, over $\mathcal R_k$ for $\mathbb E_{k,\nu}$ and over
$\mathcal R_{-k}$ for $\mathbb E^{\rm tr}_{k,\nu}$ (Appendix~\ref{app:formulas-sample}).

For each fold $k$, two fits are made on the other $K-1$ folds,
$\mathcal D_{-k}$ and $\mathcal R_{-k}$, together with the auxiliary
measurement: first the nuisances
$\hat\nu^{(-k)}$, which determine the fitted null intensity
$\hat n_k(x)=n(x\,|\,\mathrm R_{\hat\nu^{(-k)}})$, and then a departure from it,
the witness. The witness for fold $k$ is a log intensity ratio between a fitted
alternative and the fitted null:
\begin{equation}\label{eq:witness}
\phi_k(x)=\log\frac{n^{\rm alt}_k(x)}{\hat n_k(x)}\,.
\end{equation}
The held-out fold is then used only to evaluate the witness
(Section~\ref{sec:test-stat}). Cross-fitting over all $K$ folds uses every event
both to fit and to evaluate, which matters when the signal is rare.

\paragraph{Witness fit.} Any learner may produce $\phi_k$, and its form and
tuning may be chosen from the training folds and the auxiliary measurement.
The calibration of Section~\ref{sec:test-calib} conditions on the realized
witness and uses it only as a fixed function, except through the
fitting-randomness term $\hat V_{\rm fit}$, which needs the learner's
first-order response to its training events.
Section~\ref{sec:witness} details the training loss and the two
learners we use.

\paragraph{Null fit.}\label{sec:test-null}
Let $\psi$ be a vector of estimating functions, one component per nuisance.
Our nuisance fits are estimating-equation (or $M$-) estimators:
$\hat\nu^{(-k)}$ solves the estimating equations
\begin{equation}\label{eq:esteq-main}
\sum_{i\in\mathcal D_{-k}}\psi(X_i)=\mathbb E^{\rm tr}_{k,\nu}[\psi]+\Sigma_A^{-1}(\nu-\hat\nu_A)\,,
\end{equation}
which set the training sum of $\psi$ equal to its expected value under the null
model at $\nu$, corrected by the auxiliary term.
The fits differ only in the choice of $\psi$, and Appendix~\ref{app:formulas}
treats them together through this form (Eq.~\eqref{eq:esteq}).
\begin{itemize}
  \item \emph{Likelihood fit}: $\psi=s\coloneqq\partial_\nu\log n(\cdot\,|\,\mathrm R_\nu)$.
This choice of $\psi$ gives the maximum-likelihood estimate (MLE) of the nuisances:
\eqref{eq:esteq-main} then sets to zero the derivative of the extended log-likelihood of
the training events plus $\ell_A$. The exact MLE requires the score function $s$,
which is available only with oracle access. With sample access the log-ratio
$h(x;\nu)=\log[n(x\,|\,\mathrm R_\nu)/n(x\,|\,\mathrm R_0)]$, and hence the score
$s=\partial_\nu h$, is not known in closed form. Ref.~\cite{DAgnolo2022learning}
approximates $h$ by the local representation
\begin{equation}\label{eq:delta}
h(x;\nu)\simeq\nu_n+\nu_s\,\hat\delta_1(x)+\tfrac12\,\nu_s^2\,\hat\delta_2(x),
\end{equation}
whose coefficient functions $\hat\delta_{1,2}$ are learned once, on a simulation
sample disjoint from everything else, and then held fixed; we learn them in the
same way. The resulting likelihood fit, the one used in
Ref.~\cite{DAgnolo2022learning}, inherits their errors. We therefore use
the ``pushforward'' fit below.
\item \emph{Pushforward fit}: $\psi=(1,\hat\delta_1)$, the learned score at $\nu=0$. To keep the
errors of the learned score out of the estimate, we use it as a fixed estimating function and
compute its expectations by applying $T_{\nu_s}$ to the Reference events, so the fit is a
method-of-moments estimator.
With the
diagonal $\Sigma_A$ of Section~\ref{sec:setup}, \eqref{eq:esteq-main} reads
\begin{equation}\label{eq:push}
\begin{split}
N_{-k}&=\mathbb E^{\rm tr}_{k,\nu}[1]+\frac{\nu_n-\hat\nu_{A,n}}{\sigma_n^2},\\
\sum_{i\in\mathcal D_{-k}}\hat\delta_1(X_i)&=\mathbb E^{\rm tr}_{k,\nu}[\hat\delta_1]+\frac{\nu_s-\hat\nu_{A,s}}{\sigma_s^2}.
\end{split}
\end{equation}
The left-hand sides are the training count and the training sum of
$\hat\delta_1$, and the right-hand sides their expected values under the null
at $\nu$, corrected by the auxiliary term. The two sides have equal
expectation at the true $\nu^*$ for any fixed function $\hat\delta_1$, so the
estimate is consistent regardless of error in $\hat\delta_1$, which affects only
its variance.
\end{itemize}

\subsection{The held-out statistic and its centering}\label{sec:test-stat}

The witness is evaluated on the held-out fold $\mathcal D_k$, which it has not
seen.
Each event falls in $\mathcal D_k$ with probability $1/K$, so an intensity $n$
for the data is an intensity $n/K$ for the fold, with extended log-likelihood
$\ell_k(n)=\sum_{i\in\mathcal D_k}\log\{n(X_i)/K\}-\frac1K\int n$. The held-out
statistic of fold $k$ is the log-likelihood ratio of the fitted alternative to
the fitted null,
\begin{align}
\hat T_k
&=\ell_k\big(n^{\rm alt}_k\big)-\ell_k\big(\hat n_k\big)\nonumber\\
&=\Big[\sum_{i\in\mathcal D_k}\log\frac{n^{\rm alt}_k(X_i)}{K}-\frac1K\int n^{\rm alt}_k\Big]
-\Big[\sum_{i\in\mathcal D_k}\log\frac{\hat n_k(X_i)}{K}-\frac1K\int\hat n_k\Big]\nonumber\\
&=\sum_{i\in\mathcal D_k}\log\frac{n^{\rm alt}_k(X_i)}{\hat n_k(X_i)}
-\frac1K\int\big(n^{\rm alt}_k-\hat n_k\big)\nonumber\\
&=\sum_{i\in\mathcal D_k}\phi_k(X_i)-\mathbb E_{k,\hat\nu^{(-k)}}\big[e^{\phi_k}-1\big].\label{eq:Tk}
\end{align}

Held-out likelihood ratios of this kind are the basis of existing
sample-splitting tests, for example, the split likelihood-ratio test of universal
inference~\cite{Wasserman2020universal}
and the classifier-based tests for new-physics searches of
Ref.~\cite{Chakravarti2023model}. Such methods often have sub-optimal power in
practice, because they compare $\hat T_k$ with a fixed threshold calibrated by a null
centering at $0$, ignoring that
$\hat T_k$ tends to be negative under the null. To see why, suppose the held-out events
follow the fitted null; the expectation of $\hat T_k$ given the training folds is then
\begin{equation}\label{eq:mk}
m_k=-\mathbb E_{k,\hat\nu^{(-k)}}\big[e^{\phi_k}-1-\phi_k\big]\;\le\;0 ,
\end{equation}
with equality only for $\phi_k\equiv0$. A witness that has fitted noise in the
training folds predicts the held-out fold worse than the null does, so
overfitting lowers the held-out statistic instead of raising it.
We instead center $\hat T_k$ at its expectation under the true null given the
training folds,
\[m_k^*=\mathbb E_{k,\nu^*}[\phi_k]-\mathbb E_{k,\hat\nu^{(-k)}}[e^{\phi_k}-1]\,.\]
This centering is not available, because $\nu^*$ is unknown, but replacing
$\nu^*$ by its training-fold estimate $\hat\nu^{(-k)}$ gives $m_k$, and fold $k$
contributes
\begin{equation}\label{eq:Sk}
S_k=\hat T_k-m_k=\sum_{i\in\mathcal D_k}\phi_k(X_i)-\mathbb E_{k,\hat\nu^{(-k)}}[\phi_k] ,
\end{equation}
the held-out sum of the witness minus its expected value under the fitted null.
The cross-fitted test statistic is
\[S=\sum_{k=1}^KS_k\,.\]
Centering at the fitted rather than the true null changes the variance of $S$
through $\hat\nu^{(-k)}$; Section~\ref{sec:test-calib} propagates the nuisance
estimate into the variance to first order.

\subsection{Analytic calibration}\label{sec:test-calib}

Under $H_0$, $S$ has mean zero to the order relevant for the test, and the
calibration supplies its variance. We report
\begin{equation}\label{eq:Z}
Z=\frac{S}{\sqrt{\hat V}},\qquad
\hat V=\sum_{k=1}^K\big(\hat V^D_k+\hat V^R_k\big)+\hat V_A+\hat V_{\rm fit},
\end{equation}
and reject at level $\alpha$ when $Z>z_{1-\alpha}$, the one-sided Gaussian
quantile. The four variance terms correspond to the four sources of randomness in $S$:
the held-out events, the held-out Reference events, the auxiliary measurement,
and the fits themselves. Their full expressions are collected in
Appendix~\ref{app:formulas}; here we explain what each one accounts for.

\paragraph{Held-out events and the estimated nuisances.} Given the training
folds, the held-out sum in~\eqref{eq:Sk} is a sum over a Poisson process, with
variance $\mathbb E_{k,\hat\nu^{(-k)}}[\phi_k^2]$. This alone would be the right variance
if the null were known. It is not: $\hat\nu^{(-k)}$ is estimated, and an error in
it shifts the expected value subtracted in~\eqref{eq:Sk} by approximately
$g_k^\top(\hat\nu^{(-k)}-\nu^*)$, where
$g_k=\partial_\nu\mathbb E_{k,\nu}[\phi_k]$ at $\nu=\hat\nu^{(-k)}$ is the
sensitivity of the witness's null expectation to the nuisances. Because every
event takes part in $K-1$ nuisance fits, these shifts accumulate across folds
and are of the same order as $S$ itself. We account for them in the variance
by grouping each event's shifts with its own held-out fold, leading to the orthogonalized
witness (see Appendix~\ref{app:formulas-orth} for the derivation):
\begin{equation}
  \label{eq:tildephi}
  \tilde\phi_k=\phi_k-(K-1)\,g_k^\top J_k^{-1}\psi\,,
\end{equation}
where $\psi$ is the estimating function of the nuisance estimate (the score
$s$ for the likelihood fit,
$(1,\hat\delta_1)$ for the pushforward fit) and $J_k$ the
Jacobian of the corresponding estimating equations, including the auxiliary
term $\Sigma_A^{-1}$.
Then
$\hat V^D_k$ is the Poisson variance of $\tilde\phi_k$:
\begin{equation}\label{eq:VD}
\hat V^D_k=\mathbb E_{k,\hat\nu^{(-k)}}\big[\tilde\phi_k^{\,2}\big].
\end{equation}
This is the device of Neyman's $C(\alpha)$ test~\cite{Neyman:1959}, applied
here to the variance: only the part of the witness that the nuisances cannot imitate contributes to the null fluctuation.
The auxiliary measurement itself fluctuates from experiment to experiment and
moves every $\hat\nu^{(-k)}$ coherently; its contribution is
$\hat V_A=c^\top\Sigma_A^{-1}c$ with $c=\sum_kJ_k^{-\top}g_k$.

\paragraph{Held-out Reference events.} With sample access the expected value
in~\eqref{eq:Sk} is itself a sum over a finite Reference fold, which adds the
variance of that sum. Each held-out Reference event $e\in\mathcal R_k$ contributes
$\omega_k\tilde\phi_{k,e}$, where
\begin{equation}\label{eq:ve}
\tilde\phi_{k,e}=\tilde\phi_k\big(T_{\hat\nu_s^{(-k)}}X'_e\big)
\end{equation}
is the orthogonalized witness at the transformed event,
$\omega_k=N_0\,e^{\hat\nu^{(-k)}_n}/(KM_k)$ is the weight of a held-out
Reference event and $M_k$ is the number of events in $\mathcal R_k$. Given the
size of the Reference fold, the variance of their sum is estimated by the
centered sum of squares
\begin{equation}\label{eq:VR}
\hat V^R_k=\omega_k^2\sum_{e\in\mathcal R_k}\big(\tilde\phi_{k,e}-\bar{\tilde\phi}_k\big)^2 ,
\end{equation}
with $\bar{\tilde\phi}_k$ the mean of the $\tilde\phi_{k,e}$. The variance is centered, unlike the Poisson data
term, because the size of the Reference fold is fixed and carries no
information about the data.
Near the central value $\hat V^R_k$ is about $r=N_0/M$ times the data term, so at
$r=0.2$ ignoring it would inflate the false-positive rate of
a nominal $5\%$ test to about $7\%$. With oracle access $\hat V^R_k=0$.

\paragraph{Randomness of the witness fits.} The terms above treat each witness as fixed.
However, the witnesses are random and share
training events: the events held out in fold $k$ are also in the training set
of $\phi_j$, $j\ne k$, so the contributions of folds $j$ and $k$ are
correlated. When the learner is stable (its output does not move much when one
event is replaced), these correlations are negligible against the conditional
variance and $\hat V_{\rm fit}=0$ is asymptotically correct.
Under the null hypothesis in our experiments, the fitted witness is
mostly estimation noise and not stable enough, and
the cross-fold covariance of the witness must be accounted for. We estimate it by
the first-order expression derived in Appendix~\ref{app:formulas-fit},
\begin{equation}\label{eq:Vfit}
\hat V_{\rm fit}=\sum_{j\ne k}E_jE_k\,\mathrm{tr}\!\big(D_kH_k^{-1}\tilde\Omega_{kj}\,D_jH_j^{-1}\tilde\Omega_{jk}\big),
\end{equation}
where
\begin{itemize}
\item $E_k=\mathbb E_{k,\hat\nu^{(-k)}}[1]$ is the expected number of held-out
  events in fold $k$;
\item $H_k$ is the Hessian of fold $k$'s witness fit in its free directions---active atoms
in the dictionary or non-clipped parameters in the network;
\item $\tilde\Omega_{kj}$ is the second moment, under the normalized fitted
  null, of the gradients of the witnesses of folds $k$ and $j$ in their free
  directions;
\item $D_k$ is a diagonal matrix that damps the directions shared with the null
  fit (Appendix~\ref{app:impl-dict}).
\end{itemize}
No estimator of the variance of a $K$-fold
statistic computed from the fold statistics alone can be unbiased in
general~\cite{Bengio2004no}; $\hat V_{\rm fit}$ escapes this because it uses the
fitted learner's first-order response to its training events, not only the fold
statistics; it is unbiased only to first order.

\subsection{Algorithm and cost}\label{sec:test-algo}

\begin{enumerate}
\item Assign the data events, and with sample access the Reference events, to
  $K$ folds at random.
\item For each fold $k$: fit the nuisances on the other folds and the auxiliary
  measurement (Eq.~\eqref{eq:esteq-main} or Eq.~\eqref{eq:push}); fit the witness
  $\phi_k$ on the other folds; evaluate $S_k$ (Eq.~\eqref{eq:Sk}) and the fold's
  terms of the variance (Section~\ref{sec:test-calib}).
\item Combine the folds into $\hat V$ and $Z$ (Eq.~\eqref{eq:Z}) and report the
  one-sided $p$-value
  $1-\Phi(Z)$, where $\Phi$ is the standard normal cumulative distribution
  function.
\end{enumerate}
The cost is $K$ fits of the witness per data set, one per fold, with
everything else computed from quantities the fits already provide; no
pseudo-experiments are needed for the calibration. We use $K=10$ throughout.

\subsection{What is guaranteed, and where it can fail}\label{sec:test-theory}

The justification of~\eqref{eq:Z} is a central limit theorem for
cross-validated statistics~\cite{Lei2025modern}, extended to a Poisson data set
with a Reference sample and constrained nuisances. We state it informally; the
precise assumptions and a proof sketch are in Appendix~\ref{app:theory}.

\paragraph{Result.} Under $H_0$, as the expected number of events grows with
$K$ and $r$ fixed, $Z$ converges in distribution to a standard normal variable,
so the test has asymptotic size $\alpha$, provided the following conditions hold.
\begin{itemize}
\item \emph{Data} (D). The data, the Reference sample and the auxiliary measurement
  are independent, and, given their number, the Reference events are
  independent draws from the Reference model. The test conditions on the
  Reference fold sizes, which may be random but must not depend on the events
  themselves.
\item \emph{Nuisance estimate} (N). The nuisance estimates
  $\hat\nu^{(-k)}$ converge to $\nu^*$ at the usual rate, $N_0^{-1/2}$.
\item \emph{Witness} (W1, W2). The fitted departure stays bounded where the
  Reference events lie (W1). The learner is stable relative to the size of its
  own output (W2): replacing a single training event changes the witness, at a typical
  point, by an amount small compared with $\sigma_\phi/\sqrt{N_0}$, where
  $\sigma_\phi$ is the typical size of the witness (a deterministic scale, for
  example the root mean square of its standard deviation under the Reference
  model). The witness need not converge to a non-zero limit:
  $\sigma_\phi$ may shrink as the data set grows, provided the effect of a single
  event, measured relative to $\sigma_\phi$, vanishes faster than $N_0^{-1/2}$.
\end{itemize}
Under these conditions $\hat V_{\rm fit}$ is negligible and may be omitted.

Three features of the result matter for its use. First, (W2) concerns the
output of the learner, not its parameters: a clipped, penalized, early-stopped
or non-identifiable model satisfies it as long as no single event moves its fit
much, and regularization helps. Second, the held-out construction means the
test conditions on the realized witnesses: a poor witness lowers the power but
not the validity. Third, the learned function $\hat\delta_{1}$ enters our
test only as a component of the estimating function $\psi=(1,\hat\delta_1)$ in the
pushforward equations~\eqref{eq:push},
and every expectation under the null is computed by applying the exact nuisance
transformation to Reference events, so (N) holds whatever the error of
$\hat\delta_1$, which affects only the asymptotic variance of the nuisance estimate.
In contrast, in Ref.~\cite{DAgnolo2022learning}, the learned representation enters both terms of
$t=\tau-\Delta$, and it is unclear how its error affects the validity of the test.

\paragraph{The degenerate regime.} Condition (W2) can fail under the null
hypothesis whenever the learner, given no departure to learn, returns a witness
that is pure estimation noise, as small as the fitting noise itself, so the
effect of a single event is no longer small relative to it. This happened for
every witness in our experiments. The variance of $S$ then contains the cross-fold
correlations described above, and $\hat V_{\rm fit}$ is required. In this
regime the theorem does not apply and we have no limit theorem in its place.
$\hat V_{\rm fit}$ corrects the first-order bias of the conditional variance as
an estimate of $\mathrm{Var}(S)$. In the null studies of Sections~\ref{sec:1d}
and~\ref{sec:5d}, the variance-corrected statistic $Z$ is close enough to
standard normal that the false-positive rates at nominal $5\%$ and $1\%$ stay
close to nominal.

\section{Witnesses}\label{sec:witness}

The calibration of Section~\ref{sec:test} treats the witness as the output of an
arbitrary learner. This section states what that learner must
provide and describes the two we use. They were chosen to be as different as
possible: a penalized linear expansion, whose fit is a convex problem with an
exact solution, and the small neural networks of Ref.~\cite{DAgnolo2022learning},
trained by stochastic optimization for a fixed number of epochs. The same
calibration rule is applied to both.

\subsection{Requirements on the learner}\label{sec:witness-req}

A witness for fold $k$ must be a function $\phi_k(x)$ that can be evaluated at
any point, computed from the training folds and the auxiliary measurement
alone. Any choice made from those inputs is allowed, including its form,
its tuning and any modification of its output, because the calibration
conditions on the realized witness. The theory then asks for the stability and
boundedness conditions (W1) and (W2) of Section~\ref{sec:test-theory}. Both are
properties of the fitted function rather than of a parametrization, and
regularization of any kind (a penalty, weight clipping, early stopping) helps
to meet them.

Two further pieces of information are needed for the fitting-randomness term
$\hat V_{\rm fit}$ of Eq.~\eqref{eq:Vfit}, which describes to first order how
the witness responds to its training events: the directions in which the fit
is free to move, and the curvature of the training loss in those directions.
For a fit that maximizes an extended-likelihood loss both are standard
by-products; Section~\ref{sec:test-calib} and Appendix~\ref{app:impl} give them
for our witnesses. A
learner that cannot supply them can still be used when the fitting randomness is
negligible, but otherwise its null calibration
needs another estimate of $\hat V_{\rm fit}$.

All our witnesses fit the same loss, the extended log-likelihood of a tilt of
the fitted null on the training folds,
\begin{equation}\label{eq:wloss}
\mathcal L_k[f]=\sum_{i\in\mathcal D_{-k}}f(X_i)-\mathbb E^{\rm tr}_{k,\hat\nu^{(-k)}}\big[e^{f}-1\big],
\end{equation}
maximized over a class of functions $f$, possibly with a penalty, with
$\phi_k=\hat f$. This is NPLM's loss with the Reference at the fitted
nuisances in place of the central value.

\subsection{Dictionary witness}\label{sec:witness-dict}

The first witness is a penalized expansion, $f=\sum_j\beta_jA_j$, in a
dictionary of functions (atoms) $\{A_j\}$ fixed in advance, fitted by penalized
Poisson regression, a convex problem solved numerically to convergence. The
dictionary has three blocks: a tangent block, the directions in which the null
itself moves (the constant and the nuisance scores), and a smooth block for broad
distortions, both unpenalized, and a bump block of Gaussian bumps for localized
departures, with an $\ell_1$ (lasso) penalty~\cite{Tibshirani:1996lasso}.
It serves as a fast stand-in for a
black-box learner. The choice of the dictionary affects the power only, not the
calibration. We report it for the benchmark of Section~\ref{sec:1d}, where its
atoms and penalty are specified.

\subsection{Neural-network witness}\label{sec:witness-nn}

The second witness is the network of Ref.~\cite{DAgnolo2022learning}: a fully
connected network with sigmoid activations and a linear output,
trained by maximizing
Eq.~\eqref{eq:wloss} using the Adam optimizer~\cite{Kingma:2014adam} with NPLM's
settings (Appendix~\ref{app:impl-nn}) for a fixed
number of epochs, with every weight clipped to
$[-c,c]$ after each step. We use the same architectures as Ref.~\cite{DAgnolo2022learning}.

A network witness can extrapolate freely in a sparsely populated tail of the
Reference distribution, where its value on a rare held-out event can carry much
of the test statistic. When this is a concern, we can multiply the network output by
a taper $\tau_k(x)$, which equals one where the fitted null is well populated
beyond $x$ and falls to zero as $x$ increases (Appendix~\ref{app:impl-nn}). The taper depends only
on the fitted null, so the tapered output is still a valid witness. We use it
for the network of Section~\ref{sec:1d}; the network of Section~\ref{sec:5d} is
used without one, and its null studies (Section~\ref{sec:5d-valid}) show none is
needed there.

The clipping bound, the training length and, where it is used, the taper
are the only tuning choices, and are listed with each benchmark
in Sections~\ref{sec:1d} and~\ref{sec:5d}. Unlike in Ref.~\cite{DAgnolo2022learning},
they are not selected to make a statistic follow a $\chi^2$ distribution, and
much shorter training suffices.

\section{One-dimensional benchmark}\label{sec:1d}

\subsection{Setup}\label{sec:1d-setup}

The benchmark is the univariate problem of Ref.~\cite{DAgnolo2022learning},
where the Reference intensity is exponential:
\begin{equation}\label{eq:1d-ref}
n(x\,|\,\mathrm R_\nu)=N_0\,\exp\!\big[-x\,e^{-\nu_s}-\nu_s+\nu_n\big],\qquad x\ge0,
\end{equation}
with $N_0=2000$ expected events, a normalization nuisance $\nu_n$ and a scale
nuisance $\nu_s$. The three signals of Ref.~\cite{DAgnolo2022learning} are added
on top of the Reference:
\begin{enumerate}
\item[NP1:] $10$ expected events in a Gaussian peak at $x=6.4$ of width $0.16$;
\item[NP2:] $180$ expected events with the broad shape $\frac12x^2e^{-x}$;
\item[NP3:] $90$ expected events in a Gaussian peak at $x=1.6$ of width $0.16$.
\end{enumerate}
Data sets are generated with the number of
background and signal events drawn from Poisson distributions, and the auxiliary
estimates $\hat\nu_A$ are drawn from $N(\nu^*,\Sigma_A)$, where $\Sigma_A$ is diagonal
with diagonal elements $\sigma_n^2=\sigma_s^2=\sigma_A^2$.
Unlike Ref.~\cite{DAgnolo2022learning}, which
represents the Reference by a sample of $2\times10^5$ events, we use the exact
density~\eqref{eq:1d-ref}, making this a case of oracle access.

\paragraph{Dictionary witness.}
In fold $k$, let $\hat p$ be the density of the fitted null
$n(\cdot\,|\,\mathrm R_{\hat\nu^{(-k)}})/(N_0e^{\hat\nu_n^{(-k)}})$, $\hat F$ its
distribution function, $\hat q_{0.999}$ its $0.999$ quantile, and
$(\hat\mu,\hat\sigma)$ its mean and standard deviation.
For the exponential~\eqref{eq:1d-ref},
$\hat\mu=\hat\sigma=e^{\hat\nu_s^{(-k)}}$.
The dictionary witness is
\begin{equation}\label{eq:1d-dict}
f(x)=\beta_0+\beta_s\,\tilde s(x)+\sum_{l=1}^{5}\gamma_l\,P_l(x)+\sum_{j=1}^{J}\beta_j\,B_j(x),
\end{equation}
with the following atoms.
\begin{itemize}
\item $\tilde s$ is the score of the scale nuisance $\nu_s$, centered and normalized, where the
  score of $\nu_s$ is $\partial_{\nu_s}\log n(x\,|\,\mathrm R_\nu)=x\,e^{-\hat\nu_s^{(-k)}}-1$;
  together with the constant, $\tilde s$ spans the tangent block.
\item $P_1,\dots,P_5$ form an orthonormal basis of the polynomials of degree at
  most $6$ in $2\hat F(x)-1$ that are orthogonal to $1$ and to $\tilde s$, ordered so that
  $P_l$ has degree at most $l+1$.
\item $B_j$ is the Gaussian bump
  $b_j(x)=\exp\{-(x-c_j)^2/2w^2\}$ with width $w=0.2\,\hat\sigma$ and centers
  $c_j=0.1\,\hat\sigma\,(j-1)$, $j=1,\dots,J$, up to $\hat q_{0.999}$ ($J=70$),
  with its projection on $\{1,\tilde s,P_1,\dots,P_5\}$ removed, and scaled to
  $\int B_j^2\,\hat p=1$.
\end{itemize}
The three blocks are mutually orthogonal under $\hat p$, and the smooth and bump
atoms are bounded. The coefficients $\beta_0,\beta_s,\gamma_l$ are unpenalized,
and the $\beta_j$ carry an $\ell_1$ penalty of level $1.3$. This level was set once,
before any test, from null simulations: it is the rounded version of the smallest level at which, on
average, no more than $5\%$ of the bumps enter a null fit. It is used
unchanged at every $\nu^*$ and $\sigma_A$.

\paragraph{Network witness.} The network uses the architecture of Ref.~\cite{DAgnolo2022learning}:
\[f(x)=b_2+\sum_{h=1}^4W_{2,h}\,\mathrm{sigmoid}(W_{1,h}x+b_{1,h}),\]
with its eight
weights $W_{1,h},W_{2,h}$ clipped at $c=4$ and its five biases free, trained for
$3\times10^4$ epochs and tapered with $\kappa=2$ (Eq.~\eqref{eq:taper}).

\subsection{Validity}\label{sec:1d-valid}

Table~\ref{tab:1d-null} gives the false-positive rate of both witnesses at nominal
$5\%$ and $1\%$, at the central value and with each nuisance shifted by one
standard deviation, for $\sigma_A\in\{0.05,0.10,0.15\}$.
With $\sigma_A=0.05$ every rate is at or below its
nominal value within Monte Carlo error, for both witnesses and at all five
nuisance points: between $0.0375$ and $0.0475$ at $5\%$, and between $0.0025$ and
$0.0175$ at $1\%$. With $\sigma_A=0.15$ both witnesses stay within about two
standard errors of nominal at all five points, with a wider spread: between
$0.0300$ and $0.0650$ at $5\%$, and between $0.0025$ and $0.0200$ at $1\%$. The
largest deviation at $5\%$ is the network at $\sigma_A=0.10$, $0.0700$ (and
$0.0200$ at $1\%$), about two standard errors above nominal; at $\sigma_A=0.15$
the same network is at or below nominal at $5\%$ at every point, so the deviation
does not grow with $\sigma_A$. The test is slightly conservative overall: the
mean of $Z$ under the null hypothesis lies between $-0.22$ and $0.00$ in every
configuration.

\begin{table}[t]
\centering
\caption{One-dimensional benchmark: false-positive rate at nominal $5\%$ and
$1\%$, 400 data sets per entry (Monte Carlo standard errors about $0.011$ and
$0.005$). $\nu^*$ is the true value of the nuisances: $0$, or one nuisance
shifted by $\pm1$ standard deviation. At $\sigma_A=0.10$ only the central value was run.}
\label{tab:1d-null}
\begin{tabular}{llcccc}
\toprule
 & & \multicolumn{2}{c}{network} & \multicolumn{2}{c}{dictionary} \\
\cmidrule(lr){3-4}\cmidrule(lr){5-6}
$\sigma_A$ & $\nu^*$ & $5\%$ & $1\%$ & $5\%$ & $1\%$ \\
\midrule
0.05 & $0$ & 0.0375 & 0.0025 & 0.0400 & 0.0125 \\
0.05 & $+\sigma_n$ & 0.0425 & 0.0025 & 0.0475 & 0.0175 \\
0.05 & $-\sigma_n$ & 0.0400 & 0.0100 & 0.0425 & 0.0100 \\
0.05 & $+\sigma_s$ & 0.0475 & 0.0025 & 0.0475 & 0.0075 \\
0.05 & $-\sigma_s$ & 0.0475 & 0.0100 & 0.0475 & 0.0125 \\
0.10 & $0$ & 0.0700 & 0.0200 & 0.0425 & 0.0100 \\
0.15 & $0$ & 0.0500 & 0.0050 & 0.0625 & 0.0200 \\
0.15 & $+\sigma_n$ & 0.0500 & 0.0200 & 0.0300 & 0.0025 \\
0.15 & $-\sigma_n$ & 0.0325 & 0.0025 & 0.0650 & 0.0175 \\
0.15 & $+\sigma_s$ & 0.0475 & 0.0075 & 0.0450 & 0.0075 \\
0.15 & $-\sigma_s$ & 0.0450 & 0.0100 & 0.0325 & 0.0025 \\
\bottomrule
\end{tabular}

\end{table}

\begin{figure}[t]
\centering
\includegraphics[width=\textwidth]{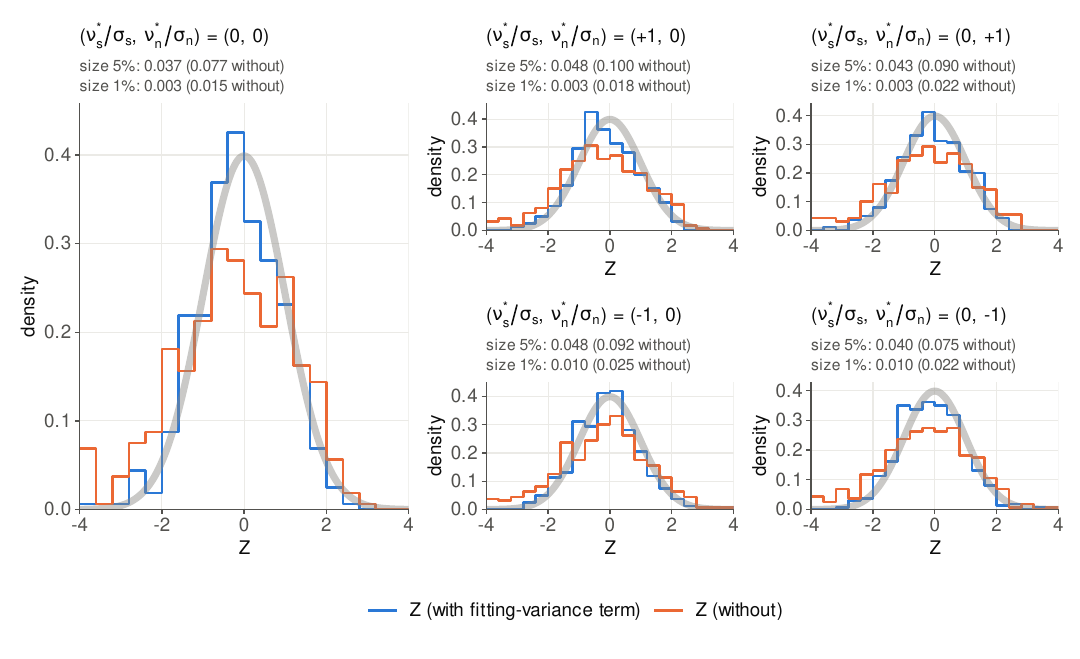}
\caption{One-dimensional benchmark, network witness, $\sigma_A=0.05$:
distribution of $Z$ under the null hypothesis at the central value (left) and
with each nuisance shifted by $\pm1$ standard deviation (right), with (blue) and
without (orange) the fitting-randomness term $\hat V_{\rm fit}$, against the
standard normal density (grey). The rates at nominal $5\%$ and $1\%$ are given
above each panel. 400 data sets per panel.}
\label{fig:1d-null}
\end{figure}

Figure~\ref{fig:1d-null} shows the distribution of $Z$ for the network at the
five nuisance points, together with the statistic obtained without the
fitting-randomness term. With
the correction the distribution follows the standard normal; without it the spread is
visibly too large and the rate at nominal $5\%$ is between $1.5$ and $2$ times
its nominal value,
which is the degenerate regime of Section~\ref{sec:test-theory} at work. The
corresponding figure for the dictionary is in Appendix~\ref{app:extra}.

Ref.~\cite{DAgnolo2022learning} validates its calibration at the same nuisance
points with a Kolmogorov--Smirnov test of the whole distribution of its statistic
on 100 data sets, whereas we measure the rejection rate in the tail directly, on
400 data sets. The two checks measure different things. The Kolmogorov--Smirnov test is
most sensitive near the center of the distribution, and with 100 data sets it can
barely see the tail, where the validity of a test is decided.

\subsection{Power}\label{sec:1d-power}

Table~\ref{tab:1d-power} gives the median $Z$, the counterpart of the median
$Z$-score $\overline Z$ of Ref.~\cite{DAgnolo2022learning}, and the power at
nominal $5\%$ and $1\%$, at the central value with $\sigma_A=0.05$. The two
witnesses have similar power on the broad NP2 and the bulk peak NP3; on the
narrow tail peak NP1 the dictionary, whose bumps have a width comparable to
that peak, is ahead. Ref.~\cite{DAgnolo2022learning} reports median $Z$-scores of similar
size for NP1 and NP3 and a larger one ($1.9\pm 0.12$) for NP2. Tighter clipping, such as $c=3$,
  would raise the median $Z$ of NP2 to about $1.27$ (Section~\ref{sec:1d-tuning}).

\begin{table}[t]
\centering
\caption{One-dimensional benchmark: median $Z$ (with its bootstrap standard
error) and power (rejection rate) at nominal $5\%$ and $1\%$, at $\nu^*=0$ and
$\sigma_A=0.05$,
400 data sets per entry.}
\label{tab:1d-power}
\begin{tabular}{lcccccc}
\toprule
 & \multicolumn{3}{c}{network} & \multicolumn{3}{c}{dictionary} \\
\cmidrule(lr){2-4}\cmidrule(lr){5-7}
signal & median $Z$ & $5\%$ & $1\%$ & median $Z$ & $5\%$ & $1\%$ \\
\midrule
NP1 & $2.41 \pm 0.15$ & 0.6450 & 0.5150 & $3.08 \pm 0.09$ & 0.7600 & 0.6650 \\
NP2 & $1.19 \pm 0.09$ & 0.3575 & 0.1525 & $1.16 \pm 0.08$ & 0.3325 & 0.1700 \\
NP3 & $3.62 \pm 0.11$ & 0.9275 & 0.8150 & $3.43 \pm 0.07$ & 0.9125 & 0.7875 \\
\bottomrule
\end{tabular}

\end{table}

\begin{figure}[t]
\centering
\includegraphics[width=\textwidth]{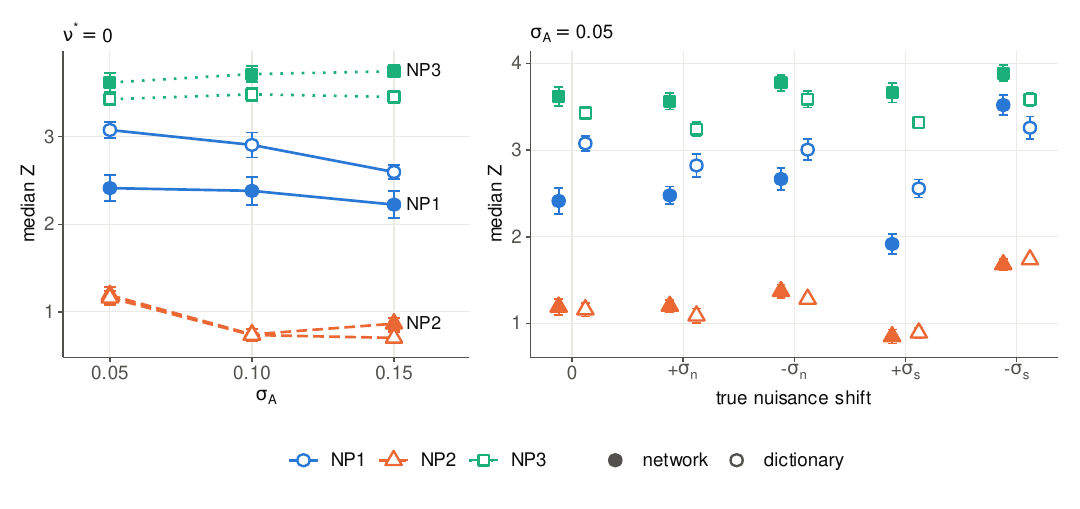}
\caption{One-dimensional benchmark: median $Z$ for NP1--NP3 (bars: bootstrap
standard error, 400 data sets per point). Left: central value $\nu^*=0$ against
the nuisance uncertainty $\sigma_A$. Right: $\sigma_A=0.05$, at the central
value and with each nuisance shifted by $\pm1$ standard deviation. Signals
are distinguished by marker shape (NP1 circle, NP2 triangle, NP3 square) and,
on the left, by line style (solid, dashed, dotted); filled markers: network,
open markers: dictionary.}
\label{fig:1d-power}
\end{figure}

Figure~\ref{fig:1d-power}, the counterpart of Fig.~8 of
Ref.~\cite{DAgnolo2022learning}, shows how the power depends on the nuisances.
As $\sigma_A$ grows from $0.05$ to $0.15$, the median $Z$ for NP2 drops from
about $1.2$ to between $0.7$ and $0.9$ for both witnesses. The median $Z$ for the
resonant NP1 drops much less, by at most $0.5$, and the one for NP3 hardly
changes. Ref.~\cite{DAgnolo2022learning} finds the
same pattern and the same reason: a change of scale can imitate a broad excess
in the tail, so a less constrained scale nuisance absorbs part of NP2, but no
change of scale or normalization produces a narrow peak. The right panel varies
the true value of the nuisances at fixed $\sigma_A$. A shift of the
normalization leaves the sensitivity nearly unchanged; a shift of the scale
moves it between $1.9$ and $3.5$ for the network (NP1), because
it changes the amount of background under the signal. Since the figure shows
median $Z$ itself rather than its ratio to a model-dependent ideal witness, this
dependence on the true scale is visible directly.

\subsection{Sensitivity to tuning}\label{sec:1d-tuning}

\begin{figure}[t]
\centering
\includegraphics[width=\textwidth]{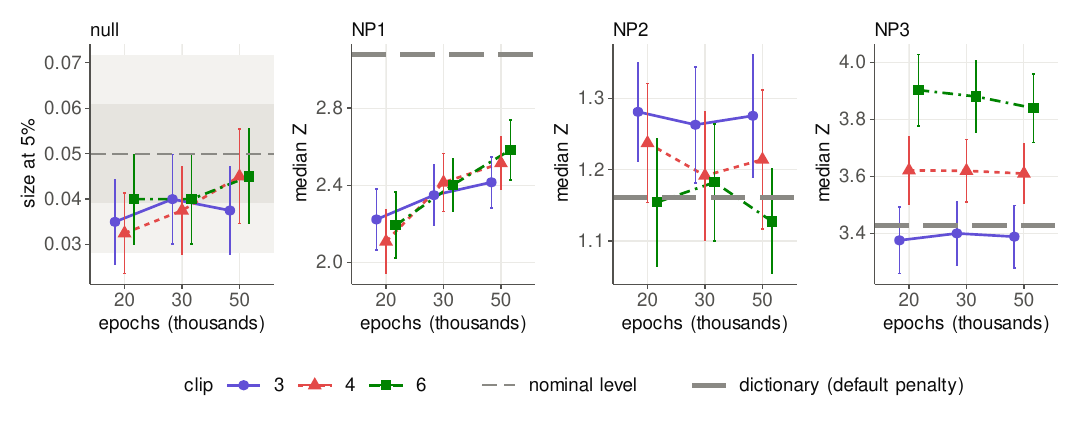}
\caption{One-dimensional benchmark, network witness, $\nu^*=0$, $\sigma_A=0.05$:
rate at nominal $5\%$ under the null hypothesis (left) and median $Z$ for
NP1--NP3 against the training length, for clipping bounds $c=3,4,6$ (each marked
by its own color, marker and line style). Bars: Monte Carlo standard errors,
400 data sets per point. Grey long-dashed lines: the nominal level (left) and the
dictionary witness at its default penalty (other panels). Shaded bands (left): the
range of $\pm1$ (darker) and $\pm2$ (lighter) Monte Carlo standard errors around
the nominal $5\%$ expected for a valid test.}
\label{fig:1d-tuning}
\end{figure}

Figure~\ref{fig:1d-tuning} varies the two tuning choices of the network, the
clipping bound $c\in\{3,4,6\}$ and the training length from $2\times10^4$ to
$5\times10^4$ epochs, with everything else fixed; Table~\ref{tab:sens-1d} in
Appendix~\ref{app:extra} lists every configuration, for both witnesses. The size is at or below
nominal in all nine configurations, at most $0.045$ at $5\%$ and $0.010$ at
$1\%$. The power moves in the directions one would expect from the shape of each
signal: a tighter bound helps the broad NP2 (median $Z$ about $1.27$ at $c=3$
against $1.15$ at $c=6$), a looser one helps the bulk peak NP3 (about $3.4$
against $3.9$), and NP1 gains from longer training. The dictionary behaves in the
same way under its penalty level $\lambda\in\{1.1,1.3,1.5\}$, run on the same
data sets: the size stays between $0.038$ and $0.045$ at $5\%$, and a heavier
penalty raises the median $Z$ of NP2 from $1.09$ to $1.21$ while leaving NP1 and
NP3 nearly unchanged. The calibration is indifferent to the tuning, and the
power responds to it as a property of the witness.

\section{Five-dimensional dilepton benchmark}\label{sec:5d}

\subsection{Data}\label{sec:5d-setup}

\paragraph{Variables.} The second benchmark is the dilepton final state of
Ref.~\cite{DAgnolo2022learning}, a case of sample access. Each event contains two
leptons, and is described by the five features
$x=(p_{T,1},p_{T,2},\eta_1,\eta_2,\Delta\phi_{12})$: the transverse momenta
$p_{T,1}\ge p_{T,2}$ of the leading and subleading lepton, that is, the components
of their momenta perpendicular to the beam axis; their pseudorapidities
$\eta_{1,2}=-\ln\tan(\theta_{1,2}/2)$, where $\theta$ is the polar angle with
respect to the beam axis; and the difference $\Delta\phi_{12}=\phi_1-\phi_2$ of
their azimuthal angles around that axis, taken in $[0,2\pi)$. The dilepton
invariant mass, which a dedicated resonance search would use, is a function of
these features and is not given to the witness separately.

\paragraph{Event selection.} We use the event selection of
Ref.~\cite{DAgnolo2022learning}, the requirements on measured quantities that
every event must meet to enter the analysis: $p_T>20$~GeV for both leptons,
$p_{T,1}<1$~TeV for the leading one, and a dilepton invariant mass above
$100$~GeV. The expected number of Standard Model events that pass it at the
central value $\nu=0$ is $N_0=8700$.

\paragraph{Nuisances.} The nuisances are the normalization, with
$\sigma_n=0.025$, and the lepton momentum scale in the electron-like regime of
Ref.~\cite{DAgnolo2022learning}, with $\sigma_s=3\times10^{-3}$. The scale acts
on simulated events through the transformation $T_{\nu_s}$ of
Section~\ref{sec:setup}: each lepton's transverse momentum is multiplied by
$e^{\nu_s}$, or by $e^{3\nu_s}$ for a lepton with $|\eta|\ge2.1$, and the
selection is then applied to the transformed event. Only $\nu_s$ changes which
events pass the selection; $\nu_n$ multiplies the expected number of events by
$e^{\nu_n}$.

\paragraph{Simulated sample.} We are given the simulated Standard Model sample
$\mathcal S$ of Ref.~\cite{Grosso2021zenodo}, $3\,735\,456$ events at the central
nuisance value, generated with requirements looser than the selection. It is
split once, at random, into two disjoint parts,
\begin{equation}\label{eq:5d-split}
\mathcal S=\mathcal S_\delta\cup\mathcal S_{\rm toy},\qquad
|\mathcal S_\delta|:|\mathcal S_{\rm toy}|=1:3 \,.
\end{equation}
$\mathcal S_\delta$ is used only to learn $\hat\delta_{1,2}$
(Section~\ref{sec:5d-est}), and $\mathcal S_{\rm toy}$ supplies the Reference
samples and the data sets. For a scale value $\nu_s$, let $\mathcal P_{\nu_s}$
be the events $X\in\mathcal S_{\rm toy}$ whose transform $T_{\nu_s}X$ passes the
selection, so that $\mathcal P_0$ are those passing at the central value. The
expected number of selected events at $\nu$ is then
\[
N_\nu=N_0\,e^{\nu_n}\,\frac{|\mathcal P_{\nu_s}|}{|\mathcal P_0|}\,.
\]
\paragraph{Reference sample and null data sets.} The two kinds of sample play
different roles. A null data set plays the role of the observed data: what an experiment
would record at the true value $\nu^*$. It is drawn in two steps: its number of
events is Poisson with mean $N_{\nu^*}$, and that many events are drawn with
replacement from $\mathcal P_{\nu_s^*}$. The Reference represents the null model
inside the test: its weighted sums, transformed to any $\nu$, estimate the
expectations $\mathbb E_{k,\nu}$ of Section~\ref{sec:test-idea}. For these sums
to estimate the expectations of the null data, the Reference is drawn in the same way,
as a null data set at the central value, with two differences: its size is fixed,
$M=5N_0=43\,500$ events from $\mathcal P_0$ (so $r=0.2$), and it also contains the
near-threshold events, in the same proportion as in $\mathcal S_{\rm toy}$
(Appendix~\ref{app:impl-samples}). Signal
data sets are built in the same way as null data sets from the signal
simulations, as described below. Each data set is drawn independently of its
Reference sample. Null studies use 400 data sets per configuration and signal
studies 160.

\paragraph{Signals.} The two signals are those of Ref.~\cite{DAgnolo2022learning}. For the $Z'$
resonance of mass $300$~GeV, the data set is a null data set plus a
sample drawn from the
$Z'$ simulation whose number of events is Poisson with mean $120$ at the
central nuisance value. For the effective-field-theory deformation with
$c_W=1$~TeV$^{-2}$, the whole data set is drawn from the EFT simulation instead
of the Standard Model one, with $1.0216\,N_0$ expected events at the central nuisance
value. In both cases the signal events are transformed by the same $T_{\nu_s^*}$ as
the Standard Model events and their expected number is scaled in the same way,
since a miscalibrated momentum scale affects every lepton. As in
Ref.~\cite{DAgnolo2022learning}, the signal data sets are generated with the scale
nuisance one standard deviation away from its central value, $\nu_s^*=+\sigma_s$.

\subsection{Estimation}\label{sec:5d-est}

\paragraph{Nuisance functions.} The functions $\hat\delta_{1,2}$ are learned on $\mathcal S_\delta$ as in
Ref.~\cite{DAgnolo2022learning}. For training points
$\nu_i\in\{\pm1.5,\pm3,\pm6,\pm10\}\times10^{-3}$ of the scale nuisance, let
$\mathcal S_1$ be the events of $\mathcal S_\delta$ that pass the selection at
$\nu=0$ and $\mathcal S_0(\nu_i)$ those that pass it after the transformation
$T_{\nu_i}$.
Two networks
$\delta_1,\delta_2$ (five hidden layers of ten units each) minimize
\begin{equation}\label{eq:delta-loss}
\begin{gathered}
\sum_i\Big[\sum_{x\in\mathcal S_0(\nu_i)}c_i(x)^2+\sum_{x\in\mathcal S_1}\big(1-c_i(x)\big)^2\Big],\\
c_i(x)=\frac{1}{1+\exp\{\nu_i\,\delta_1(x)+\tfrac12\nu_i^2\,\delta_2(x)\}},
\end{gathered}
\end{equation}
whose minimizer makes $\nu_i\delta_1+\frac12\nu_i^2\delta_2$ the log-ratio of the
shifted to the nominal event density. A fifth of the events are held out to select, among $2000$
training epochs, the networks with the lowest validation loss.
Of the two functions, our test uses only $\hat\delta_1$, in the estimating
function $\psi=(1,\hat\delta_1)$ of the pushforward fit, Eq.~\eqref{eq:push},
and of the orthogonalized witness, Eq.~\eqref{eq:tildephi}.

\paragraph{Witness.} The witness is the network of Section~\ref{sec:witness-nn} with the
architecture of Ref.~\cite{DAgnolo2022learning}, three hidden layers of five units (96 parameters), all
parameters clipped at $c=0.7$, trained for $3000$ epochs, with $K=10$ folds; the
features are standardized as in Ref.~\cite{DAgnolo2021learning}. In this
higher-dimensional setting the dictionary witness was much weaker than the
network in exploratory studies, and we do not report it.

\subsection{Validity}\label{sec:5d-valid}

\begin{figure}[t]
\centering
\includegraphics[width=\textwidth]{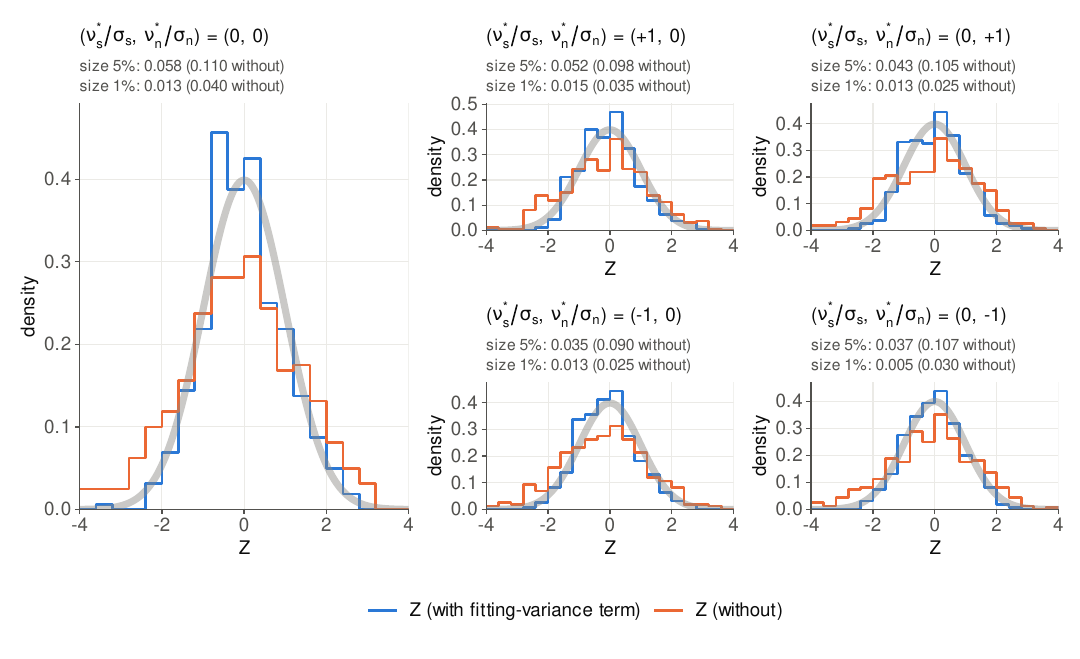}
\caption{Dilepton benchmark, network witness: distribution of $Z$ under the null
hypothesis at the central value (left) and with each nuisance shifted by one
standard deviation (right), with (blue) and without (orange) the
fitting-randomness term, against the standard normal density (grey). 400 data
sets per panel.}
\label{fig:5d-null}
\end{figure}

Table~\ref{tab:5d-null} and Fig.~\ref{fig:5d-null} give the null distribution of
$Z$ at the central value and with each nuisance shifted by one standard
deviation, the five points at which Ref.~\cite{DAgnolo2022learning} validates its
own calibration. The rate at nominal $5\%$ lies between $0.0350$ and $0.0575$,
and at nominal $1\%$ between $0.0050$ and $0.0150$, all within about one and a
half Monte Carlo standard errors of nominal. The mean of $Z$ is within $0.08$ of
zero and its standard deviation between $0.93$ and $0.97$, so the test is, if
anything, slightly conservative. Without the fitting-randomness term the rate at
nominal $5\%$ rises to between $0.09$ and $0.11$ (Fig.~\ref{fig:5d-null}): the
degenerate regime is present here as in one dimension, and the correction
$\hat V_{\rm fit}$, including its Reference component, removes it.

\begin{table}[t]
\centering
\caption{Dilepton benchmark, network witness: mean and standard deviation of
$Z$ under the null hypothesis, and its rate at nominal $5\%$ and $1\%$, at the
central value and with each nuisance shifted by one standard deviation. 400 data
sets per row (Monte Carlo standard errors about $0.011$ and $0.005$).}
\label{tab:5d-null}
\begin{tabular}{lcccc}
\toprule
$\nu^*$ & mean of $Z$ & s.d.\ of $Z$ & rate at $5\%$ & rate at $1\%$ \\
\midrule
$0$ & $-0.012$ & 0.965 & 0.0575 & 0.0125 \\
$+\sigma_s$ & $-0.012$ & 0.928 & 0.0525 & 0.0150 \\
$-\sigma_s$ & $-0.075$ & 0.928 & 0.0350 & 0.0125 \\
$+\sigma_n$ & $0.016$ & 0.942 & 0.0425 & 0.0125 \\
$-\sigma_n$ & $0.005$ & 0.949 & 0.0375 & 0.0050 \\
\bottomrule
\end{tabular}

\end{table}

\subsection{Power: \texorpdfstring{$Z'$}{Z'} and EFT}\label{sec:5d-power}

\begin{figure}[t]
\centering
\includegraphics[width=0.62\textwidth]{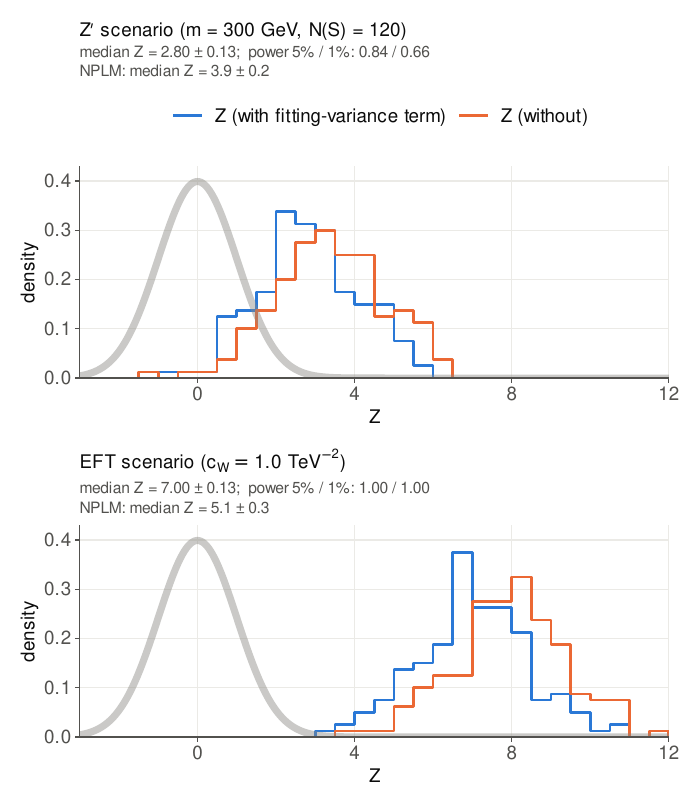}
\caption{Dilepton benchmark, network witness: distribution of $Z$ under the
$Z'$ (top) and EFT (bottom) signals at $\nu_s^*=+\sigma_s$, with (blue) and
without (orange) the fitting-randomness term; grey: the standard normal density
expected under the null hypothesis. The median $Z$ and power, and the median
$Z$-score of Ref.~\cite{DAgnolo2022learning} for the same signal, are given above
each panel. 160 data sets per panel.}
\label{fig:5d-power}
\end{figure}

Figure~\ref{fig:5d-power} shows the distribution of $Z$ under the two signals,
and Table~\ref{tab:5d-nplm} compares its median with the corresponding numbers in
Ref.~\cite{DAgnolo2022learning}.
Our network is less
sensitive than NPLM to the narrow $Z'$ resonance (median $Z$ $2.80$ against
NPLM's median $Z$-score $\overline Z=3.9$, both converted from the test's own
$p$-value) and more sensitive to the broad EFT deformation ($7.00$
against $5.1$). This is the pattern of Section~\ref{sec:1d}: the calibration is
the same for both signals, and the power reflects what the witness, here a
small network trained for $3000$ epochs with tight clipping, captures well. A
smooth, broad deformation is easy for it to fit; a narrow resonance in a sparse
region of five dimensions is harder for it to resolve.

\begin{table}[t]
\centering
\caption{Dilepton benchmark: sensitivity of the network witness compared with
Ref.~\cite{DAgnolo2022learning}, electron-like regime, $\nu_s^*=+\sigma_s$. Our
median $Z$ (bootstrap standard error) and power from 160 data sets; NPLM's median
$Z$-score $\overline Z$ as printed in Ref.~\cite{DAgnolo2022learning} (100 data
sets).}
\label{tab:5d-nplm}
\begin{tabular}{lcccc}
\toprule
 & \multicolumn{3}{c}{this work (NN witness)} & NPLM \\
signal & median $Z$ & power 5\% & power 1\% & $\overline{Z}$ \\
\midrule
$Z'$ & $2.80 \pm 0.13$ & 0.84 & 0.66 & $3.9 \pm 0.2$ \\
EFT & $7.00 \pm 0.13$ & 1.00 & 1.00 & $5.1 \pm 0.3$ \\
\bottomrule
\end{tabular}

\end{table}

\subsection{Sensitivity to tuning and computational cost}\label{sec:5d-tuning}

\begin{figure}[t]
\centering
\includegraphics[width=0.85\textwidth]{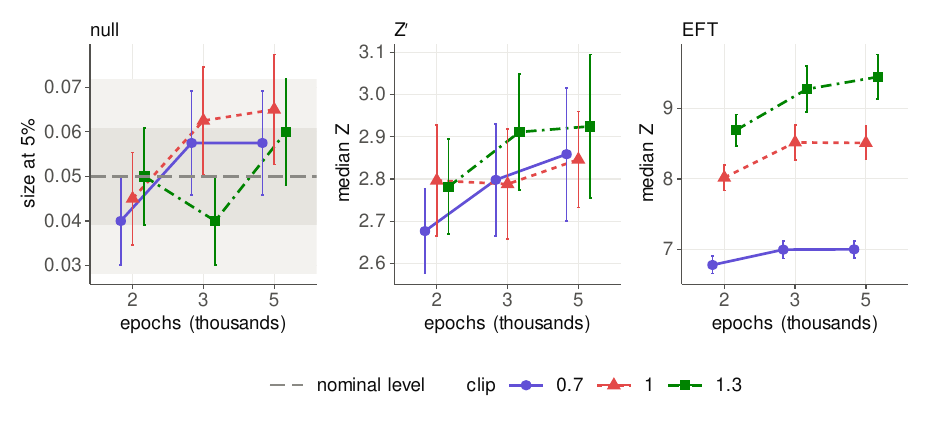}
\caption{Dilepton benchmark, network witness: rate at nominal $5\%$ under the
null hypothesis (left; 400 data sets per point) and median $Z$ for $Z'$ and EFT
(160 data sets per point) against the training length, for clipping bounds
$c=0.7,1.0,1.3$, each marked by its own color, marker and line style. Bars: Monte
Carlo standard errors. Shaded bands (left): the range of $\pm1$ (darker) and
$\pm2$ (lighter) Monte Carlo standard errors around the nominal $5\%$ expected
for a valid test.}
\label{fig:5d-tuning}
\end{figure}

Figure~\ref{fig:5d-tuning} varies the clipping bound, $c\in\{0.7,1.0,1.3\}$, and
the training length, from $2000$ to $5000$ epochs; Table~\ref{tab:sens-5d} in
Appendix~\ref{app:extra} lists every configuration. The rate at nominal $5\%$
stays between $0.040$ and $0.065$ in all nine configurations, within about one
and a half Monte Carlo standard errors of nominal. The power again responds to
the witness: the median $Z$ for $Z'$ hardly moves (between $2.68$ and $2.92$),
while that for the EFT rises with the clipping bound, from about $7.0$ at
$c=0.7$ to between $8.7$ and $9.5$ at $c=1.3$. Our default $c=0.7$ is therefore
conservative for the EFT; a looser bound would increase its sensitivity further
without affecting the calibration.

The network can be trained briefly, without having to reach a $\chi^2$ regime:
$3000$ epochs here, against $5\times10^5$ or more in
Ref.~\cite{DAgnolo2022learning}, where a single training takes a few hours of CPU
time. On one GPU, training the ten networks of one dilepton data set takes about
half a minute.

\section{Discussion}\label{sec:discussion}

We have presented a goodness-of-fit test for new-physics searches in which a
flexible learner fits the departure from the Reference model on part of the data
and is evaluated on the rest, cross-fitted so that every event is used for both,
and calibrated analytically, with explicit terms for the estimated nuisances, the
finite Reference sample and the randomness of the fits. On the benchmarks of
Ref.~\cite{DAgnolo2022learning}, the false-positive rate stayed close to nominal in
every configuration we checked. In both benchmarks this held at the central value
and with each nuisance shifted by one standard deviation. In the one-dimensional
benchmark it also held for nuisance uncertainties up to $\sigma_A=0.15$. No pseudo-experiment entered the
calibration; the pseudo-experiments in this paper serve only to check it.

\paragraph{What the calibration buys.} The central consequence is that the choice
of learner is freed from the calibration.
Our calibration
conditions on the realized fits, so a
learner is admissible as it is, within the limits discussed below, and its
tuning can be chosen for power alone.
Two consequences follow. The power depends, as it should, on how well the learner
captures the signal, and the best tuning differs from signal to signal: in one
dimension a tighter clipping bound or a heavier penalty helped the broad NP2 and a
looser bound the bulk peak NP3, while in five dimensions a looser bound helped the
EFT deformation. Throughout, the calibration
did not change. And the fits are short, because nothing
requires them to reach a $\chi^2$ regime.

\paragraph{Predictability, computability and stability.} The construction of our
predictive goodness-of-fit test can be
read through the predictability, computability and stability (PCS) framework of
veridical data science~\cite{Yu2020veridical,Yu2024veridical}. Predictability is
the basis of detection: a fitted departure counts as evidence only to the extent
that it improves the prediction of future events. Computability shows
in the cost: regularization and early stopping are used as they are, rather than
tuned until the statistic reaches a $\chi^2$ regime, and no pseudo-experiments or
resampling enter the calibration. Stability~\cite{Yu2013stability} is the starting
point of the calibration, through the central limit theorem for cross-validated
statistics, but not all of it: under the null hypothesis our learners were
not stable enough for that theory alone, and it is completed by the correction for
the fitting randomness.

\paragraph{Limitations and outlook.} We list the points on which the results
above rest on approximation or on evidence that is less than complete, and where
they leave room for further work.
\begin{itemize}
\item \emph{The degenerate regime.} Under the null hypothesis our learners were
  not stable enough for the theorem of Section~\ref{sec:test-theory} to apply
  directly, and the calibration relied on the cross-fold correction
  $\hat V_{\rm fit}$. The correction is a first-order approximation.
  No limit theorem covers
  this regime, and the normal reference rests on the null studies of
  Sections~\ref{sec:1d} and~\ref{sec:5d}. For the network, its
  linearization treats each parameter as either free or held at the clipping
  bound and carries one regularization constant fixed beforehand
  (Appendix~\ref{app:impl-nn}). The linearization is adequate
  at the configurations we use and can
  under-count the fitting variance when many parameters reach the bound.
\item \emph{Witness construction.} Power depends on how well the witness
  captures the signal, and we did not engineer the witness for power.
  Held-out evaluation and short training have a price on some signals: at the
  operating points we chose, our network is less sensitive than NPLM to the $Z'$
  resonance, and both of our witnesses are less sensitive than NPLM to the broad
  one-dimensional excess NP2. Conversely, the width of the dictionary's bumps was
  set at the scale of the narrow benchmark signals, which may favor its power there.
  Because the calibration
  conditions on whatever witness each fold produces, the framework leaves room
  for engineering the witness further, which we have not explored: larger
  networks or longer training; a dictionary adapted to the signals sought; tuning
  made data-driven, with the clipping bound or penalty selected within each fold
  by validation on its training events alone; or other learners, for example the
  kernel methods already used with NPLM~\cite{Letizia2022learning}. NPLM's own
  architectures can be used unchanged, as here, so that the two approaches can
  share a learner and differ only in how it is calibrated. Finally, the toy-free
  calibration makes it practical to scan many learners or
many regions of a data set, provided the resulting multiplicity of tests is
accounted for.
\item \emph{Two nuisances.} Both benchmarks have a normalization and a scale
  nuisance. The construction is written for any finite set of nuisances,
  but realistic analyses
  have many more, and the cost of the pushforward fit and of the variance terms
  with many nuisances has not been studied here.
\end{itemize}

\section*{Data and code availability}
The five-dimensional dilepton samples are publicly available~\cite{Grosso2021zenodo};
the one-dimensional data sets are simulated. The code that reproduces all
simulation results, tables and figures of this paper is publicly available at
\url{https://github.com/jinglei/predictive-gof}.

\appendix
\section{Calibration formulas in full}\label{app:formulas}

This appendix collects the formulas behind Eq.~\eqref{eq:Z} as they are computed,
and derives the orthogonalized witness~\eqref{eq:tildephi} and the fitting-randomness term~\eqref{eq:Vfit},
using first-order approximations. Sections~\ref{app:formulas-esteq}--\ref{app:formulas-fit}
assume oracle access; Section~\ref{app:formulas-sample} lists what changes with
sample access.

\subsection{Estimating equations}\label{app:formulas-esteq}

Every term needs expectations of functions of $x$ under the fitted null of a
fold. We write them for a function $u$ as
\begin{equation}\label{eq:Ek}
\mathbb E_{k,\hat\nu^{(-k)}}[u]=\frac1K\int\hat n_k\,u ,\qquad
\mathbb E^{\rm tr}_{k,\hat\nu^{(-k)}}[u]=\frac{K-1}K\int\hat n_k\,u ,
\end{equation}
the expected sum of $u$ over the held-out and the training events of fold $k$,
and $E_k=\mathbb E_{k,\hat\nu^{(-k)}}[1]$ for the expected number of held-out events under the fitted null $\hat n_k$.
$\mathbb E_{k,\nu}$ and $\mathbb E^{\rm tr}_{k,\nu}$ denote the corresponding expectations with
$\hat n_k$ replaced by $n(\cdot\,|\,\mathrm R_\nu)$. The corresponding empirical sums over
the data events are $\hat{\mathbb E}_k[u]=\sum_{i\in\mathcal D_k}u(X_i)$ and
$\hat{\mathbb E}^{\rm tr}_k[u]=\sum_{i\in\mathcal D_{-k}}u(X_i)$.

Both nuisance fits of Section~\ref{sec:test-null} solve estimating equations of the form
\begin{equation}\label{eq:esteq}
F_{k,\psi}(\nu)\coloneqq\hat{\mathbb E}^{\rm tr}_k[\psi]-\mathbb E^{\rm tr}_{k,\nu}[\psi]-\Sigma_A^{-1}(\nu-\hat\nu_A)=0 ,
\end{equation}
where $\psi$ is a vector of estimating functions, which may or may not depend on $\nu$, one per nuisance. We write
\begin{equation}\label{eq:Jdef}
J_k=\Big\{\partial_\nu\,\mathbb E^{\rm tr}_{k,\nu}[\psi_\nu]-\mathbb E^{\rm tr}_{k,\hat\nu^{(-k)}}[\partial_\nu\psi_\nu]+\Sigma_A^{-1}\Big\}_{\nu=\hat\nu^{(-k)}}\approx -\frac{\partial F_{k,\psi}}{\partial\nu}\Big|_{\nu=\hat\nu^{(-k)}}
\end{equation}
for the approximate sign-reversed Jacobian of $F_{k,\psi}$ with the data sum
replaced by the expectation under the fitted null; the
subscript in $\psi_\nu$ makes explicit any dependence of $\psi$ on $\nu$.
Differentiating the intensity, $\partial_\nu n(\cdot\,|\,\mathrm R_\nu)=n(\cdot\,|\,\mathrm R_\nu)\,s$
with $s=\partial_\nu\log n(\cdot\,|\,\mathrm R_\nu)$ the score,
the first term
is $\mathbb E^{\rm tr}_{k,\hat\nu^{(-k)}}[\psi s^\top]+\mathbb E^{\rm tr}_{k,\hat\nu^{(-k)}}[\partial_\nu\psi_\nu]$,
the terms in $\partial_\nu\psi_\nu$ cancel, and
\begin{equation}\label{eq:Jgen}
J_k=\mathbb E^{\rm tr}_{k,\hat\nu^{(-k)}}\big[\psi\,s^\top\big]+\Sigma_A^{-1},
\end{equation}
with rows indexing the equations and columns the nuisances. The two fits differ
only in $\psi$. With oracle access the likelihood fit has
$\psi=s$, and
\begin{equation}\label{eq:Jlik}
J_k=\mathbb E^{\rm tr}_{k,\hat\nu^{(-k)}}\big[s\,s^\top\big]+\Sigma_A^{-1},
\end{equation}
the Fisher information of the training fit, auxiliary term included, so
that $J_k^{-1}$ is the asymptotic variance of $\hat\nu^{(-k)}$. This $J_k$ is
symmetric, so $J_k^{-\top}=J_k^{-1}$ in what follows; the transposes are kept
because they matter for the pushforward fit used with sample access
(Section~\ref{app:formulas-sample}).

\subsection{Derivation of the orthogonalized witness}\label{app:formulas-orth}

In the notation of Section~\ref{app:formulas-esteq}, Eq.~\eqref{eq:Sk} reads
$S_k=\hat{\mathbb E}_k[\phi_k]-\mathbb E_{k,\hat\nu^{(-k)}}[\phi_k]$. For now we treat
the witnesses as fixed and defer the study of their own randomness to
Section~\ref{app:formulas-fit}. $S_k$ then depends on the data through the
held-out sum and, through the nuisance estimate $\hat\nu^{(-k)}$, on the
training events and the auxiliary measurement. Expanding around $\nu^*$ to first order,
\begin{equation}\label{eq:Sklin}
S_k=\Big\{\hat{\mathbb E}_k[\phi_k]-\mathbb E_{k,\nu^*}[\phi_k]\Big\}-g_k^\top\big(\hat\nu^{(-k)}-\nu^*\big)+\dots,\end{equation}
where
\[
g_k=\partial_\nu\,\mathbb E_{k,\nu}[\phi_k]\big|_{\nu=\hat\nu^{(-k)}}=\mathbb E_{k,\hat\nu^{(-k)}}[\phi_k\,s]\,.
\]

The first term in \eqref{eq:Sklin} is the held-out sum centered at its null expectation. For
the second term, $F_{k,\psi}(\hat\nu^{(-k)})=0$ and
$-\partial F_{k,\psi}/\partial\nu\approx J_k$, so linearizing~\eqref{eq:esteq}
around $\nu^*$ gives
\begin{equation}\label{eq:nulin}
\hat\nu^{(-k)}-\nu^*=J_k^{-1}F_{k,\psi}(\nu^*)+\dots
=J_k^{-1}\Big[\hat{\mathbb E}^{\rm tr}_k[\psi]-\mathbb E^{\rm tr}_{k,\nu^*}[\psi]+\Delta_A\Big]+\dots,
\end{equation}
with $\Delta_A=\Sigma_A^{-1}(\hat\nu_A-\nu^*)$.
The training sum of fold $k$ is the sum of the held-out sums of the other
folds, $\hat{\mathbb E}^{\rm tr}_k=\sum_{j\ne k}\hat{\mathbb E}_j$, and
likewise $\mathbb E^{\rm tr}_{k,\nu^*}=\sum_{j\ne k}\mathbb E_{j,\nu^*}$,
since at a common $\nu$ the expectation does not depend on the fold.
Substituting~\eqref{eq:nulin} into~\eqref{eq:Sklin}, summing over $k$ and
regrouping the training terms by the fold $j$ in which their events are held
out,
\begin{equation}\label{eq:Slin}
S=\sum_{j=1}^K\Big\{\hat{\mathbb E}_j[\tilde\phi_j]-\mathbb E_{j,\nu^*}[\tilde\phi_j]\Big\}-c^\top\Delta_A+\dots,
\end{equation}
with
\begin{equation}\label{eq:phitilde-sum}
\tilde\phi_j=\phi_j-\Big(\sum_{k\ne j}J_k^{-\top}g_k\Big)^{\!\top}\psi ,
\qquad c=\sum_{k=1}^KJ_k^{-\top}g_k .
\end{equation}
To leading order $J_k^{-\top}g_k$ does not depend on $k$, since the folds'
fits differ only by the exchange of a fraction $1/K$ of the events, so
$\sum_{k\ne j}J_k^{-\top}g_k\approx(K-1)J_j^{-\top}g_j=b_j$ and
$\tilde\phi_j=\phi_j-b_j^\top\psi$ is the orthogonalized witness of
Eq.~\eqref{eq:tildephi}, the form that is computed.

Table~\ref{tab:contrib} reads~\eqref{eq:Slin} one unit of noise at a time. A
data event $X_i$ held out in fold $j$ enters through its evaluation
$\phi_j(X_i)$ in $S_j$ and through the nuisance fits of the other $K-1$ folds,
each of which it shifts by $J_k^{-1}\psi(X_i)$; together these give
$\tilde\phi_j(X_i)$. The auxiliary measurement is never held out, so it has no
evaluation; it enters the fits of all $K$ folds, and its shifts, driven by one
random vector, add coherently to $-c^\top\Delta_A$.

\begin{table}[ht]
\centering
\caption{Contribution to $S$ of one unit of each source of noise, with the
witnesses fixed. A data event belongs to fold $j$ and enters the
nuisance fits of the $K-1$ folds $k\neq j$; the auxiliary measurement enters
the fits of all $K$ folds. The middle columns are per fold $k$ whose fit the
unit enters; the last column sums the evaluation and all such shifts, as
in~\eqref{eq:Slin}. With sample access a third row is added
(Section~\ref{app:formulas-sample}).}
\label{tab:contrib}
\small
\setlength{\tabcolsep}{4pt}
\begin{tabular}{lcccc}
\toprule
source & evaluation & shift of $\hat\nu^{(-k)}$ & shift of $S_k$ & total \\
\midrule
data event $X_i$ & $\phi_j(X_i)$ & $J_k^{-1}\psi(X_i)$ & $-g_k^\top J_k^{-1}\psi(X_i)$ & $\tilde\phi_j(X_i)$ \\
auxiliary measurement & --- & $J_k^{-1}\Delta_A$ & $-g_k^\top J_k^{-1}\Delta_A$ & $-c^\top\Delta_A$ \\
\bottomrule
\end{tabular}
\end{table}

\paragraph{Variance with fitted witnesses.} Write
\[U_j\coloneqq\hat{\mathbb E}_j[\tilde\phi_j]-\mathbb E_{j,\nu^*}[\tilde\phi_j]\]
 so that \eqref{eq:Slin} becomes
\[ S=\sum_jU_j-c^\top\Delta_A+\dots\,.
\]
$U_j$ collects the
contributions of the events in $\mathcal D_j$ through their evaluation and the
other folds' nuisance fits. Let
$\mathcal F_{-j}$ denote everything except the events of fold $j$: the other
folds and the auxiliary measurement. The witness $\phi_j$, its sensitivity
$g_j$, $J_j$ and hence the computed $\tilde\phi_j=\phi_j-b_j^\top\psi$ are
functions of $\mathcal F_{-j}$, and the events in $\mathcal D_j$ are independent of
$\mathcal F_{-j}$, so $\mathbb E[U_j\,|\,\mathcal F_{-j}]=0$. Treating $c$
as fixed, the covariances of $U_j$ with $c^\top\Delta_A$ therefore vanish, and
\begin{equation}\label{eq:VarS}
\mathrm{Var}(S)\approx\sum_j\mathrm{Var}(U_j)+\mathrm{Var}(c^\top\Delta_A)+\sum_{j\ne k}\mathrm{Cov}(U_j,U_k)\,.
\end{equation}
By the law of total variance, applied to each term with its own conditioning,
$\mathrm{Var}(U_j)=\mathbb E\big[\mathrm{Var}(U_j\,|\,\mathcal F_{-j})\big]$,
whose plug-in estimate is $\hat V^D_j$. The auxiliary term is
$\mathrm{Var}(c^\top\Delta_A)=c^\top\Sigma_A^{-1}c=V_A$, since
$\mathrm{Var}(\Delta_A)=\Sigma_A^{-1}$.
With fitted witnesses the cross terms do not vanish, because $\tilde\phi_j$
is trained on the events of fold $k$ and $\tilde\phi_k$ on those of fold $j$;
they are the fitting-randomness term addressed in Section~\ref{app:formulas-fit}.

\subsection{The fitting-randomness term}\label{app:formulas-fit}

For a fixed fold $k$, the witness is $\phi_k=f_{\hat\theta_k}$, where
$\hat\theta_k$ maximizes the training objective~\eqref{eq:wloss} over a
parameter vector $\theta$, possibly with a penalty. In the notation of
Section~\ref{app:formulas-esteq} the objective reads
\begin{equation}\label{eq:wlossE}
\mathcal L_k(\theta)=\hat{\mathbb E}^{\rm tr}_k[f_\theta]-\mathbb E^{\rm tr}_{k,\hat\nu^{(-k)}}\big[e^{f_\theta}-1\big],
\end{equation}
and $\hat\theta_k$ solves the estimating equations
\begin{equation}\label{eq:wscore}
G_k(\theta)\coloneqq\partial_\theta\mathcal L_k(\theta)=\hat{\mathbb E}^{\rm tr}_k[a_\theta]-\mathbb E^{\rm tr}_{k,\hat\nu^{(-k)}}\big[e^{f_\theta}a_\theta\big]=0 ,
\end{equation}
where $a_\theta=\partial_\theta f_\theta$ is restricted to the free directions
of fold $k$ (the active atoms of the dictionary, the free parameters of the
network; Appendix~\ref{app:impl}), and $a_k=a_{\hat\theta_k}$. As
in~\eqref{eq:esteq}, $G_k$ is a data sum minus an expectation, here under the
tilted null $e^{f_\theta}\hat n_k$. We write
\begin{align}
H_k&=\Big\{\partial_\theta\,\mathbb E^{\rm tr}_{k,\hat\nu^{(-k)}}\big[e^{f_\theta}a_\theta\big]-\mathbb E^{\rm tr}_{k,\hat\nu^{(-k)}}\big[e^{f_\theta}\partial_\theta a_\theta\big]\Big\}_{\theta=\hat\theta_k}\nonumber\\
&\approx-\frac{\partial G_k}{\partial\theta}\Big|_{\theta=\hat\theta_k}\label{eq:Hdef}
\end{align}
for the approximate sign-reversed Jacobian of $G_k$ with the data sum
$\hat{\mathbb E}^{\rm tr}_k[\partial_\theta a_\theta]$ replaced by its
expectation under the fitted tilted null. This is the witness counterpart of $J_k$ in \eqref{eq:Jdef}. A similar cancellation leads to
\begin{equation}\label{eq:Hk}
H_k=\mathbb E^{\rm tr}_{k,\hat\nu^{(-k)}}\big[e^{\phi_k}a_ka_k^\top\big]\,.
\end{equation}
For the dictionary
$a_\theta$ is the vector of active atoms and does not depend on $\theta$, so
there is no term to cancel and $H_k$ is the Hessian of the penalized Poisson
regression on its active columns. For the network the cancellation is the
Gauss--Newton approximation, and $H_k$ is the Gauss--Newton
matrix~\eqref{eq:GN}, with its ridge.

\paragraph{First-order covariance.} Under $H_0$ the
fitted witness is close to a parameter $\theta_0$ with
$f_{\theta_0}\equiv0$. Linearizing~\eqref{eq:wscore} around $\theta_0$, as
Eq.~\eqref{eq:nulin} linearizes~\eqref{eq:esteq},
$\hat\theta_k-\theta_0=H_k^{-1}G_k(\theta_0)+\dots$. We center at the true
value, so that each sum depends on the events of one fold only:
\begin{equation}\label{eq:xi}
\xi_j(u)=\hat{\mathbb E}_j[u]-\mathbb E_{j,\nu^*}[u]
\end{equation}
is the centered sum of a function $u$ over the events of fold $j$; $\xi_j(a_k)$
is that of fold $k$'s free directions,
and
$G_k(\theta_0)=\sum_{j\ne k}\xi_j(a_k)-\mathbb E^{\rm tr}_{k,\hat\nu^{(-k)}}[a_k\,s^\top]\big(\hat\nu^{(-k)}-\nu^*\big)+\dots$.
The second term is the part of the training fluctuation absorbed by the
nuisance fit, against which the witness is fitted. We set it aside here and
account for it in the computation by a damping of the nuisance directions
(Appendix~\ref{app:impl-dict}); then
\begin{equation}\label{eq:thetalin}
\hat\theta_k-\theta_0\approx H_k^{-1}\sum_{j\ne k}\xi_j(a_k) .
\end{equation}
The witness is, to first order, $\phi_k=a_k^\top(\hat\theta_k-\theta_0)$.
Starting from the definition of $U_k$ in Section~\ref{app:formulas-orth},
with $b_k$ treated as fixed given $\mathcal F_{-k}$,
\begin{align}
U_k&=\xi_k(\tilde\phi_k)\nonumber\\
&=\xi_k(\phi_k)-b_k^\top\xi_k(\psi)\nonumber\\
&=\xi_k(a_k)^\top(\hat\theta_k-\theta_0)-b_k^\top\xi_k(\psi)+\dots\nonumber\\
&=\xi_k(a_k)^\top H_k^{-1}\sum_{j\ne k}\xi_j(a_k)-b_k^\top\xi_k(\psi)+\dots\label{eq:Skfit}
\end{align}
The first line is the definition of $U_k$ in the notation of~\eqref{eq:xi},
the second splits $\tilde\phi_k=\phi_k-b_k^\top\psi$, using that $\xi_k$ is
linear, the third substitutes the linearized witness
$\phi_k=a_k^\top(\hat\theta_k-\theta_0)$, and the last
uses~\eqref{eq:thetalin}.
The first term carries the fitting noise: the held-out events of fold $k$
paired with the training events. For $j\ne k$, $U_j$ and $U_k$ share two such
pairs: the events of fold $k$ enter $U_k$ as held-out events and $U_j$ as
training events, and vice versa for fold $j$. Since the folds are independent
Poisson processes and each centered sum has mean zero, only these pairs
contribute:
\begin{align}
\mathrm{Cov}(U_j,U_k)
&=\mathrm{Cov}\big(\xi_j(a_j)^\top H_j^{-1}\xi_k(a_j),\ \xi_k(a_k)^\top H_k^{-1}\xi_j(a_k)\big)\nonumber\\
&=\mathrm{tr}\Big(H_j^{-1}\,\mathbb E\big[\xi_k(a_j)\,\xi_k(a_k)^\top\big]\,H_k^{-1}\,\mathbb E\big[\xi_j(a_k)\,\xi_j(a_j)^\top\big]\Big)\nonumber\\
&=E_jE_k\,\mathrm{tr}\big(H_k^{-1}\Omega_{kj}H_j^{-1}\Omega_{jk}\big).\label{eq:covUjUk}
\end{align}
The second line writes the product of the two scalars as
$x^\top Ay\,z^\top Bw=\mathrm{tr}(A\,yz^\top B\,wx^\top)$ and takes the
expectation fold by fold: $\xi_k(a_j)$ and $\xi_k(a_k)$ depend on the events
of fold $k$, $\xi_j(a_k)$ and $\xi_j(a_j)$ on those of fold $j$, and the folds
are independent. The last line uses Campbell's theorem,
$\mathbb E\big[\xi_k(a_j)\,\xi_k(a_k)^\top\big]=\mathbb E_{k,\nu^*}[a_ja_k^\top]=E_k\Omega_{jk}$,
and the cyclic property of the trace, where, in the ideal case,
\begin{equation}\label{eq:Qstar}
\Omega_{kj}=\int a_k\,a_j^\top\,dQ^*,\qquad Q^*=\frac{n(\cdot\,|\,\mathrm R_{\nu^*})}{\int n(\cdot\,|\,\mathrm R_{\nu^*})},
\end{equation}
is the second moment of the free directions of folds $k$ and $j$ under the
normalized true null; how it is computed is described in
Appendix~\ref{app:impl-vfit}. With $H_k^{-1}$ replaced by $D_kH_k^{-1}$,
the damping of Appendix~\ref{app:impl-dict}, this is the summand of
Eq.~\eqref{eq:Vfit}.
When all folds share the
same $p$ unpenalized directions, orthonormal under $Q^*$, the sum over
$j\ne k$ equals the conditional variance $\sum_k\hat V^D_k$ to leading order: in
the linear case the fitting randomness doubles the variance. For such a
linear witness with known nuisances, standard linear algebra shows that, given
the first-order expansion of the fit, this correction is exact.

\subsection{Sample access: what changes}\label{app:formulas-sample}

With sample access the Reference model is known only through a simulated
sample, and every expectation $\mathbb E_{k,\nu}$ or $\mathbb E^{\rm tr}_{k,\nu}$
in Sections~\ref{app:formulas-esteq}--\ref{app:formulas-fit} that is computed,
rather than only written, is replaced by its Reference estimate
$\bar{\mathbb E}_{k,\nu}$ or $\bar{\mathbb E}^{\rm tr}_{k,\nu}$. The
difference $\bar{\mathbb E}-\mathbb E$ is the Reference noise. It adds a
third source of noise to Table~\ref{tab:contrib}, the term $\hat V^R_k$ to the
variance, and a Reference part to the overlaps of $\hat V_{\rm fit}$.

\paragraph{Reference expectations.} With $T_{\nu_s}$ the nuisance transformation
followed by the event selection and $M_k$, $M_{-k}$ the numbers of held-out and
training Reference events that pass the selection at $\nu=0$,
\begin{equation}\label{eq:Eksample}
\bar{\mathbb E}_{k,\nu}[u]=\frac{N_0\,e^{\nu_n}}{K\,M_k}\sum_{e\in\mathcal R_k}\!\mathbf 1[T_{\nu_s}X'_e\ \text{passes}]\,u(T_{\nu_s}X'_e),
\end{equation}
and $\bar{\mathbb E}^{\rm tr}_{k,\nu}$ is the same sum over
$\mathcal R_{-k}$ with weight $(K-1)N_0e^{\nu_n}/(K\,M_{-k})$. At the fitted
value $\nu=\hat\nu^{(-k)}$ the held-out weight is
$\omega_k=N_0\,e^{\hat\nu_n^{(-k)}}/(K\,M_k)$, with $\hat\nu_n^{(-k)}$ the normalization
component of $\hat\nu^{(-k)}$. Both weights
are close to $r\,e^{\hat\nu_n^{(-k)}}$, since each held-out fold holds about a fraction
$1/K$ of the Reference sample. The Reference folds are drawn with fixed
sizes (Appendix~\ref{app:impl-vr}). The sensitivity $g_k$, which carries the
nuisance estimate into the variance through the orthogonalization $b_k$, is
computed from these sums: its entry for $\nu_n$ is
$g_{k,n}=\bar{\mathbb E}_{k,\hat\nu^{(-k)}}[\phi_k]$, because the weight is
proportional to $e^{\nu_n}$, and its entry for $\nu_s$, like the
$\nu_s$-derivatives in $J_k$, is a central difference of these sums, with the
selection repeated at each shifted value (Appendix~\ref{app:impl-num}).

\paragraph{Pushforward nuisance fit.} The default fit~\eqref{eq:push} has
$\psi=(1,\hat\delta_1)$, a fixed function. Because the intensity is proportional
to $e^{\nu_n}$, its two training expectations factor as
$\bar{\mathbb E}^{\rm tr}_{k,\nu}[1]=e^{\nu_n}C_{0,k}(\nu_s)$ and
$\bar{\mathbb E}^{\rm tr}_{k,\nu}[\hat\delta_1]=e^{\nu_n}C_{1,k}(\nu_s)$. The
entries of~\eqref{eq:Jgen} are their derivatives: in $\nu_n$ the expectations
themselves, and in
$\nu_s$ the derivatives $C_{0,k}'$ and $C_{1,k}'$, computed on the training Reference,
\begin{equation}\label{eq:Jpush}
J_k=e^{\hat\nu_n^{(-k)}}\begin{pmatrix}C_{0,k} & C_{0,k}'\\ C_{1,k} & C_{1,k}'\end{pmatrix}+\Sigma_A^{-1}.
\end{equation}

\paragraph{Reference noise in $U_j$.} The subtracted expectations are now
Reference sums, and the centered sum~\eqref{eq:xi} becomes
\[
\xi_j(u)=\hat{\mathbb E}_j[u]-\bar{\mathbb E}_{j,\nu^*}[u]
=\Big\{\hat{\mathbb E}_j[u]-\mathbb E_{j,\nu^*}[u]\Big\}
-\Big\{\bar{\mathbb E}_{j,\nu^*}[u]-\mathbb E_{j,\nu^*}[u]\Big\},
\]
the data noise minus the Reference noise of fold $j$, both functions of the
data and Reference events of fold $j$ only. The training Reference events
enter the nuisance fits through $\bar{\mathbb E}^{\rm tr}_{k,\nu}[\psi]$, and
$\bar{\mathbb E}^{\rm tr}_{k,\nu}=\sum_{j\ne k}\bar{\mathbb E}_{j,\nu}$ up to
the imbalance of the fold sizes, so the regrouping of
Section~\ref{app:formulas-orth} goes through unchanged:
$S=\sum_jU_j-c^\top\Delta_A+\dots$ with $U_j=\xi_j(\tilde\phi_j)$, and
$\mathcal F_{-j}$ now excludes the Reference events of fold $j$ as well. A
Reference event $X'_e$ of fold $j$ therefore acts as a data event with the
opposite sign and weight $\omega_j$: every entry of its row in
Table~\ref{tab:contrib} is $-\omega_j$ times the corresponding entry of the
data row, with $\phi_j(e)$ and $\psi(e)$ evaluated at the transformed event
$T_{\hat\nu_s^{(-j)}}X'_e$ and set to zero if it fails the selection, and its
total contribution is
\begin{equation}\label{eq:refrow}
-\omega_j\,\tilde\phi_j(e).
\end{equation}
The data and Reference parts of $\xi_j$ are
independent, so $\mathrm{Var}(U_j\,|\,\mathcal F_{-j})$ in~\eqref{eq:VarS} is
the sum of the data part, estimated by $\hat V^D_j$, and the Reference part,
the variance of the sum of the totals~\eqref{eq:refrow} over $\mathcal R_j$,
estimated by the centered $\hat V^R_j$ of Eq.~\eqref{eq:VR}
(Appendix~\ref{app:impl-vr}); it is approximately
$\omega_j^2\sum_e\tilde\phi_{j,e}^{\,2}=\omega_j\,\bar{\mathbb E}_{j,\hat\nu^{(-j)}}[\tilde\phi_j^{\,2}]$,
about $r$ times the data part.

\paragraph{Fitting randomness.} The Reference events of fold $j$ are also in
the training samples of the witnesses $\phi_k$, $k\ne j$, so the cross terms
$\mathrm{Cov}(U_j,U_k)$ are derived as in Section~\ref{app:formulas-fit}, with
the updated definition of $\xi_k$. The data and Reference parts
of $\xi_k$ are independent, and the data part gives
$\mathbb E_{k,\nu^*}[a_ja_k^\top]=E_k\Omega_{jk}$ as before. The Reference part
is $\omega_k$ times a sum over the $M_k$ Reference events of fold $k$, a fixed
number of draws from the normalized null $Q^*$. The
covariance of such a sum is $M_k$ times the covariance of a single draw,
$\Omega_{jk}-\mu_j\mu_k^\top$, rather than the raw second moment $\Omega_{jk}$
that a Poisson count gives: with a fixed number of draws there is no
fluctuation of the count to add $\mu_j\mu_k^\top$. With $\omega_kM_k=E_k$, the expected number of held-out
events, and $\omega_k\approx r$ near the central value,
\begin{align}
\mathbb E\big[\xi_k(a_j)\,\xi_k(a_k)^\top\big]
&=\mathbb E_{k,\nu^*}[a_ja_k^\top]+\omega_k^2M_k\,\mathrm{Cov}_{Q^*}(a_j,a_k)\nonumber\\
&=E_k\Omega_{jk}+\omega_kE_k\big(\Omega_{jk}-\mu_j\mu_k^\top\big)\nonumber\\
&\approx E_k\tilde\Omega_{jk}\nonumber
\end{align}
with
\begin{equation}\label{eq:Mtilde}
\tilde\Omega_{kj}=\Omega_{kj}+r\big(\Omega_{kj}-\mu_k\mu_j^\top\big),\qquad\mu_k=\int a_k\,dQ^* .
\end{equation}
Here $Q^*$ is the normalized true null of Eq.~\eqref{eq:Qstar}. The factor $r$ comes from the weights: a Reference fold has about $1/r$
times as many events as the data fold, each weighted by $r$, so its sum has
$(1/r)\,r^2=r$ times the variance of the data sum. Had the Reference
fold been drawn as a Poisson process, its moment would be uncentered like the
data part, and $\tilde\Omega=(1+r)\Omega$.

\subsection{Summary}\label{app:formulas-summary}

For each fold the computation returns $S_k$, $g_k$, $J_k$, $\tilde\phi_k$ on the
held-out data and the fitted null, $\hat V^D_k$,
$\hat V^R_k$ and, for $\hat V_{\rm fit}$, the free
directions on the common measure $\hat Q$ (Appendix~\ref{app:impl-vfit})
with $H_k$ and $D_k$. The statistic is
\begin{equation}\label{eq:Zfull}
\begin{split}
Z&=\frac{\sum_kS_k}{\sqrt{\hat V}},\\
\hat V&=\sum_k\big(\mathbb E_{k,\hat\nu^{(-k)}}[\tilde\phi_k^{\,2}]+\hat V^R_k\big)+c^\top\Sigma_A^{-1}c\\
&\quad+\sum_{j\ne k}E_jE_k\,\mathrm{tr}\big(D_kH_k^{-1}\tilde\Omega_{kj}D_jH_j^{-1}\tilde\Omega_{jk}\big).
\end{split}
\end{equation}
With oracle access the expectations are exact, $\hat V^R_k=0$ and
$\tilde\Omega_{kj}=\Omega_{kj}$; with sample access every computed $\mathbb E$ is the
Reference estimate $\bar{\mathbb E}$ of Section~\ref{app:formulas-sample}.
The numerator uses the raw witness $\phi_k$ and the variance the orthogonalized
$\tilde\phi_k$ (Section~\ref{sec:test-calib}).
\section{Statements of the limit theorems and proof sketches}\label{app:theory}

We state the result of Section~\ref{sec:test-theory} in the form used here and
sketch its proof. Precise statements, the proofs, and the extension to the
degenerate regime will be given in a companion paper in statistics.

\paragraph{Setting and scale.} The expected number of events $N_0$ grows, with
$K$, $r=N_0/M$ and $\Sigma_A N_0$ fixed. Let $\sigma_\phi=[\mathbb E\,\mathrm{Var}_{Q^*}(\phi_k)]^{1/2}$
be the typical size of the witness, where $\mathrm{Var}_{Q^*}(\phi_k)$ is the conditional variance of $\phi_k$
under $Q^\ast$ in \eqref{eq:Qstar}.
Let $\sigma_S^2=\sum_k(V^D_k+V^R_k)+V_A$, with $V^D_k$ and $V^R_k$ the conditional
variances given the training folds that $\hat V^D_k$ and $\hat V^R_k$ estimate
(Appendices~\ref{app:impl-vd} and~\ref{app:formulas-sample}).
The conditions of Section~\ref{sec:test-theory} read:
\begin{description}
\item[(D)] $\mathcal D$ is a Poisson process with intensity
  $n(\cdot\mid\mathrm R_{\nu^*})$; given the stratum sizes (Appendix~\ref{app:impl-samples}), the Reference
  events are independent draws from the Reference model; $\hat\nu_A\sim
  N(\nu^*,\Sigma_A)$; the three are independent, and the fold assignment is
  random within strata.
\item[(N)] Each $\hat\nu^{(-k)}$ admits the linearization~\eqref{eq:nulin}
  with remainder $o_P(N_0^{-1/2})$, and $\nu\mapsto\mathbb E_{k,\nu}[\phi_k]$ is
  twice differentiable near $\nu^*$ with second derivative $O_P(N_0\sigma_\phi)$.
\item[(W1)] With probability tending to one,
  $\sup_x|\phi_k(x)|\le B\sigma_\phi$ on the support of the Reference events,
  for a constant $B$.
\item[(W2)] Replacing one training data event by an independent copy changes
  $\phi_k$ at an independent point by $o(\sigma_\phi/\sqrt{N_0})$ in $L_2$;
  replacing one training Reference event, whose weight in the loss is about
  $r$, changes it by $o(\sqrt r\,\sigma_\phi/\sqrt{N_0})$.
\end{description}

\paragraph{Theorem.} Under $H_0$ and (D), (N), (W1), (W2),
\[
\frac{S}{\sqrt{\sum_k\big(\hat V^D_k+\hat V^R_k\big)+\hat V_A}}\ \rightsquigarrow\ N(0,1).
\]
The cross-fold covariance that $\hat V_{\rm fit}$ estimates is
$o(\sigma_S^2)$, so $Z$ of Eq.~\eqref{eq:Z} has the same limit whenever
$\hat V_{\rm fit}/\sigma_S^2\to0$ in probability. With oracle access the
Reference terms are absent. No condition involves $\hat\delta_{1,2}$: the
nuisance transformation is applied exactly to Reference events, so the learned
functions affect only the precision of $\hat\nu^{(-k)}$.

\paragraph{Proof sketch.} The argument is the random-centering central limit
theorem for cross-validation~\cite{Lei2025modern}, applied to two input types
and an auxiliary measurement.
\begin{enumerate}
\item \emph{Plug-in.} Expanding $S_k$ around $\nu^*$ and
  substituting~\eqref{eq:nulin} gives
  $S=\sum_jU_j-c^\top\Delta_A+R$, with $U_j=\xi_j(\tilde\phi_j)$ as in
  Appendix~\ref{app:formulas-orth}: $\xi_j$ is the centered sum~\eqref{eq:xi}
  over the data events of fold $j$, minus its Reference counterpart with
  sample access (Appendix~\ref{app:formulas-sample}), and $\Delta_A$ is as
  in~\eqref{eq:nulin}. By (N) and (W1) the remainder $R$, second order in
  $\hat\nu^{(-k)}-\nu^*$, is $o_P(\sigma_S)$.
\item \emph{Stabilization.} Let $\bar\phi_j(x)=\mathbb E[\tilde\phi_j(x)]$,
  the expectation over the training events of fold $j$ (data and Reference)
  and $\hat\nu_A$ at fixed $x$. By the Efron--Stein inequality, (W2), summed
  over the training events of both types, gives
  $\tilde\phi_j=\bar\phi_j+o_P(\sigma_\phi)$ in $L_2(Q^*)$.
  Replacing $\tilde\phi_j$ by $\bar\phi_j$ therefore changes $S$ by $o_P(\sigma_S)$,
  and the dependence of $S_j$ on the evaluation events of $S_k$, $j\neq k$,
  contributes $o(\sigma_S^2)$ to $\mathrm{Var}(S)$.
\item \emph{Central limit theorem.} The leading term is a sum of independent
  contributions: a compound-Poisson sum over the data events, whose variance is
  uncentered; a sum over the Reference events, centered within strata; and the
  Gaussian $-c^\top\Delta_A$. Its variance is $\sigma_S^2$, and (W1) gives the
  Lindeberg condition.
\item \emph{Studentization.} Given the training folds, $\hat V^D_k$ and
  $\hat V^R_k$ are unbiased for $V^D_k$ and $V^R_k$, and by (W1), (W2) the sum
  is ratio-consistent for $\sigma_S^2-V_A$. Slutsky's lemma concludes.
\end{enumerate}

Condition (W2) can fail under $H_0$, when the witness is pure estimation noise
(Section~\ref{sec:test-theory}). Step~2 then breaks down, the cross-fold
covariance is no longer negligible, and the correction of
Appendix~\ref{app:formulas-fit} is needed.
\section{Implementation details}\label{app:impl}

This appendix describes how the quantities of Appendix~\ref{app:formulas} are
computed: the Reference sample with sample access, the expectations and their
nuisance derivatives, the nuisance fits and the learned functions
$\hat\delta_{1,2}$ they use with sample access, the variance terms, and the
fitting-randomness term for each witness.

\subsection{The Reference sample with sample access}\label{app:impl-samples}

In the benchmark of Section~\ref{sec:5d} the selection depends on the scale
nuisance, so a Reference made only of the events that pass at $\nu=0$ would
lose, after a transformation $T_{\nu_s}$, the events that the shift brings into the
selection. The Reference is therefore drawn before the selection. The scale
shifts considered lie in $\mathcal V_s=[-\bar\nu_s,\bar\nu_s]$,
$\bar\nu_s=0.065/3\approx0.022$ (about $7\sigma_s$), which contains every value
used in the fits and in the data sets. Let $\mathcal B$ be the events of
$\mathcal S_{\rm toy}$ that fail the selection at $\nu=0$ but pass it with its
$p_T$ and invariant-mass thresholds loosened by a factor $e^{0.065}$. Since no
momentum moves by more than this factor for $\nu_s\in\mathcal V_s$, $\mathcal B$
contains every event that such a shift can bring into the selection,
\[
\bigcup_{\nu_s\in\mathcal V_s\setminus\{0\}}\mathcal P_{\nu_s}\setminus\mathcal P_0\subseteq\mathcal B ,
\]
with $\mathcal P_{\nu_s}$ as in Section~\ref{sec:5d-setup}. The Reference
consists of $M$ events drawn with replacement from $\mathcal P_0$ and, in
addition, $n_{\mathcal B}=M\,|\mathcal B|/|\mathcal P_0|$ events drawn with replacement
from $\mathcal B$, so that the two kinds appear in the same proportion as in
$\mathcal S_{\rm toy}$. Its weights are normalized by $M$, the events that pass
at $\nu=0$ (Eq.~\eqref{eq:Eksample}), and an event contributes to an
expectation at $\nu$ only if $T_{\nu_s}$ brings it into the selection. Drawing the
second stratum from all of $\mathcal S_{\rm toy}\setminus\mathcal P_0$ at the
same rate would give the same expectations: restricting it to $\mathcal B$ only
omits events that pass the selection at no $\nu_s\in\mathcal V_s$ and therefore
contribute zero, which saves transforming them at every $\nu_s$. The two strata
are kept separate when the Reference is split into folds
(Appendix~\ref{app:impl-vr}).

\subsection{Expectations and nuisance derivatives}\label{app:impl-num}
With oracle access the integrals of Eq.~\eqref{eq:Ek} are computed by the
trapezoid rule on a uniform grid that covers the support of the fitted null up
to its $1-10^{-9}$ quantile, with $16\,001$ nodes for the dictionary witness and
$4\,001$ for the network.

With sample access the expectations are the Reference sums~\eqref{eq:Eksample}.
Their derivatives in $\nu_n$ are the sums themselves; in $\nu_s$
they are central differences,
\begin{equation}\label{eq:gs}
g_{k,s}=\frac{\bar{\mathbb E}_{k,(\hat\nu_n^{(-k)},\hat\nu_s^{(-k)}+\varepsilon)}[\phi_k]-\bar{\mathbb E}_{k,(\hat\nu_n^{(-k)},\hat\nu_s^{(-k)}-\varepsilon)}[\phi_k]}{2\varepsilon},\qquad\varepsilon=10^{-3},
\end{equation}
where the held-out Reference events are transformed to the shifted value and
the selection is repeated, so that the events that cross a selection threshold
are counted. The derivatives $C_{0,k}'$ and $C_{1,k}'$ in Eq.~\eqref{eq:Jpush} are
computed in the same way on the training Reference.

\subsection{Nuisance fits}\label{app:impl-nufit}

\paragraph{Likelihood fit.} With oracle access the likelihood fit, Eq.~\eqref{eq:esteq-main} with
$\psi=s$, is solved by Newton's method, started at the unconstrained estimate
($\Sigma_A^{-1}=0$) and iterated until the residual of the estimating
equations is below $10^{-10}$. In the benchmark of Section~\ref{sec:1d} the
normalization and scale equations separate into two concave one-dimensional
problems, and the iteration is monotone after at most one overshoot.

\paragraph{Pushforward fit.} The count equation of~\eqref{eq:push} is strictly
decreasing in $\nu_n$ at fixed $\nu_s$, and is solved for $\nu_n$ by Newton's
method to $10^{-12}$. Substituting this profiled value $\nu_n(\nu_s)$ leaves
one equation in the scale,
\[
F_k(\nu_s)=\sum_{i\in\mathcal D_{-k}}\hat\delta_1(X_i)-e^{\nu_n(\nu_s)}C_{1,k}(\nu_s)-\frac{\nu_s-\hat\nu_{A,s}}{\sigma_s^2}=0 ,
\]
with $C_{1,k}$ as in Appendix~\ref{app:formulas-sample}.
$C_{1,k}$ is a sum over the training Reference with the selection repeated at each
$\nu_s$, so it jumps whenever a Reference event crosses a selection threshold;
the root is therefore found by Brent's method, which needs no derivative, to an
absolute tolerance of $10^{-6}$ in $\nu_s$. The search is confined to
$|\nu_s|\le10^{-2}$, the range of scale shifts on which $\hat\delta_{1,2}$ are
trained and within which the Reference retains every event the shift can bring
into the selection (Appendix~\ref{app:impl-samples}). The full-data fit is solved
first on this interval. Each fold's fit starts from a bracket of half-width
$10^{-3}$ around the full-data estimate, widened by a factor of four until
$F_k$ changes sign or the bracket reaches the interval. If $F_k$ does not
change sign on the interval, the estimate is set to the bound and the fit is
flagged. In the studies of Section~\ref{sec:5d} ($2320$ data sets, $23\,200$ fold
fits) no fit reached the bound; the largest $|\hat\nu_s^{(-k)}|$ was $0.0076$. %

\paragraph{The learned functions $\hat\delta_{1,2}$.} The samples $\mathcal S_1$
and $\mathcal S_0(\nu_i)$ of the loss~\eqref{eq:delta-loss} are built from the
same events of $\mathcal S_\delta$, before and after $T_{\nu_i}$. The inputs are preprocessed as in
Ref.~\cite{DAgnolo2021learning}: the pseudorapidities are standardized and the
positive features ($p_{T,1}$, $p_{T,2}$, $\Delta\phi_{12}$) divided by their mean,
with statistics from the training events of $\mathcal S_1$. Both networks have
ReLU hidden units and a linear output, are initialized by the Glorot-uniform
rule, and are trained jointly, full batch, with Adam at the settings of
Appendix~\ref{app:impl-nn}. The validation loss on the held-out fifth (split
by event, so that an event and its shifted copies fall on the same side) is
evaluated every ten epochs. In the test the networks are evaluated by a forward
pass that reproduces the trainer's output to within $10^{-8}$.

\subsection{Computing \texorpdfstring{$\hat V_k^D$}{V\_k\^{}D}}\label{app:impl-vd}

$\hat V^D_k$ is the variance of the sum, over the events held out in fold
$k$, of each event's total contribution $\tilde\phi_k(X_i)$ in
Table~\ref{tab:contrib}. Given the training folds, the held-out events
approximately form a Poisson process with intensity $\hat n_k/K$: their
number is random as well as their locations. By Campbell's
theorem~\cite{Kingman:1993poisson} the variance of
$\hat{\mathbb E}_k[\tilde\phi_k]$ is the uncentered second moment
$\hat V^D_k=\mathbb E_{k,\hat\nu^{(-k)}}[\tilde\phi_k^{\,2}]$; the fluctuation of the count is
part of the null fluctuation of $S$.

\subsection{Computing \texorpdfstring{$\hat V_k^R$}{V\_k\^{}R}}\label{app:impl-vr}

The Reference sample has two strata (Appendix~\ref{app:impl-samples}): the
events that pass the selection at $\nu=0$ and the near-threshold events that
fail it, which $T_{\hat\nu_s^{(-k)}}$ can move into the acceptance; an event that
still fails at $\hat\nu^{(-k)}$ has $\tilde\phi_{k,e}=0$. Let $\mathcal P_{0,k}$ and
$\mathcal B_{0,k}$ be the held-out Reference events of fold $k$ in the two
strata, with sizes $M_k=|\mathcal P_{0,k}|$ and $n_{\mathcal B,k}=|\mathcal B_{0,k}|$.
The folds are drawn separately within the strata, so $M_k$ and $n_{\mathcal B,k}$ are
fixed, the weight $\omega_k$ is normalized by $M_k$, and only which events
were drawn is random. Within a stratum the held-out events are then a simple
random sample, and the variance of their sum is the centered within-stratum
sum of squares,
\begin{equation}\label{eq:VRstrat}
\hat V^R_k=\omega_k^2\Big[\sum_{e\in\mathcal P_{0,k}}\big(\tilde\phi_{k,e}-\bar{\tilde\phi}^{\mathcal P}_k\big)^2
+\sum_{e\in\mathcal B_{0,k}}\big(\tilde\phi_{k,e}-\bar{\tilde\phi}^{\mathcal B}_k\big)^2\Big] ,
\end{equation}
with $\bar{\tilde\phi}^{\mathcal P}_k$ and $\bar{\tilde\phi}^{\mathcal B}_k$ the means over
$\mathcal P_{0,k}$ and $\mathcal B_{0,k}$. Centering within strata rather than
over the whole fold matters: $\mathcal B_{0,k}$ consists mostly of events with
$\tilde\phi_{k,e}=0$, and the difference between the stratum means, which does not
fluctuate when the stratum sizes are fixed, would otherwise be counted as
variance. With a single stratum, Eq.~\eqref{eq:VRstrat} reduces
to~\eqref{eq:VR}. Had the
fold been drawn as a Poisson process, Campbell's theorem would apply as for
the data term and give the uncentered $\omega_k^2\sum_e\tilde\phi_{k,e}^{\,2}$; the two differ by
$\omega_k^2\big[M_k(\bar{\tilde\phi}^{\mathcal P}_k)^2+n_{\mathcal B,k}(\bar{\tilde\phi}^{\mathcal B}_k)^2\big]$.

\subsection{Computing \texorpdfstring{$\hat V_{\rm fit}$}{V\_fit}}\label{app:impl-vfit}

\paragraph{Poisson moments.} Under $H_0$ the data events form a Poisson
process with intensity $n(\cdot\,|\,\mathrm R_{\nu^*})$, and the random
assignment of events to folds splits it into $K$ independent Poisson processes,
each with intensity $n(\cdot\,|\,\mathrm R_{\nu^*})/K$. For vector functions
$b$ and $c$, Campbell's theorem~\cite{Kingman:1993poisson} gives the
covariance of their sums over the events of fold $l$ as an uncentered second
moment,
\begin{equation}\label{eq:campbell-cov}
\mathrm{Cov}\big(\hat{\mathbb E}_l[b],\hat{\mathbb E}_l[c]\big)=\mathbb E_{l,\nu^*}\big[b\,c^\top\big],
\end{equation}
of which Appendix~\ref{app:impl-vd} is the case $b=c=\tilde\phi_l$; sums over
different folds are independent. With the free directions held fixed, the
centered sums~\eqref{eq:xi} therefore have
$\mathbb E\big[\xi_l(a_j)\xi_l(a_k)^\top\big]=\mathbb E_{l,\nu^*}[a_ja_k^\top]$
for every fold $l$, and are independent across $l$.

\paragraph{Pairing.} Plugging~\eqref{eq:Skfit} into Eq.~\eqref{eq:covUjUk}, only the shared pairs
contribute:
$\xi_j(a_j)^\top H_j^{-1}\xi_l(a_j)\,\xi_k(a_k)^\top H_k^{-1}\xi_m(a_k)$ with
 $l=k$ and $m=j$, because the folds are independent and each $\xi$ has
mean zero.

\paragraph{Common measure and overlaps.} Writing
$\mathbb E_{l,\nu^*}[\,\cdot\,]=E_l\int\cdot\;dQ_l$, with $Q_l$ the normalized
null of fold $l$, Eq.~\eqref{eq:covUjUk} needs the second moments of the free
directions of two folds under the null of a third. The folds' fitted nulls
differ only through their nuisance estimates, at relative order
$N_0^{-1/2}$, so the computation replaces all of them, and $Q^*$, by one
common probability measure $\hat Q$. With oracle access, $\hat Q$ is the
normalized fitted null of the first fold, evaluated on its integration grid.
With sample access, it is the normalized null at the full-data nuisance fit,
represented by the Reference sample. The
computed overlaps are
\begin{equation}\label{eq:Mkj}
\Omega_{kj}=\int a_k\,a_j^\top\,d\hat Q ,
\end{equation}
uncentered for the same reason as $\hat V^D_k$, and with the damping of
Appendix~\ref{app:impl-dict} the summand of $\hat V_{\rm fit}$ is
$E_jE_k\,\mathrm{tr}\big(D_kH_k^{-1}\Omega_{kj}D_jH_j^{-1}\Omega_{jk}\big)$. The
curvature is computed on the same measure, rescaled to the training exposure
of fold $k$: $H_k=E^{\rm tr}_k\int e^{\phi_k}a_ka_k^\top\,d\hat Q$, with
$E^{\rm tr}_k=\mathbb E^{\rm tr}_{k,\hat\nu^{(-k)}}[1]$ the expected number
of training events. With sample access the overlaps acquire the Reference
part of Eq.~\eqref{eq:Mtilde}.

\subsection{Damping}\label{app:impl-dict}

The linearized witness~\eqref{eq:thetalin} sets aside the
second term of $G_k(\theta_0)$, the part of the training fluctuation absorbed by
the nuisance fit. Along a direction in which the null itself can move, a
training fluctuation is shared between the null and the witness: the nuisance
fit takes a fraction $\rho$ of it and the witness keeps the rest. The auxiliary
measurement limits how far the null can follow the data, so the share $1-\rho$
left to the witness lies between $0$, for an unconstrained nuisance, which
absorbs the direction entirely, and $1$, for a nuisance pinned by the auxiliary
measurement. The damping scales the fitting randomness of such directions by
this share.

Formally, substituting~\eqref{eq:nulin} into the second term of
$G_k(\theta_0)$, its data part replaces the training sums of $a_k$ by those of
the residualized directions
\[
\tilde a_k=a_k-\mathbb E^{\rm tr}_{k,\hat\nu^{(-k)}}[a_k\,s^\top]\,J_k^{-1}\psi ,
\]
the free directions with the part absorbed by the nuisance fit removed. Two
atoms of the dictionary are themselves nuisance scores, and we call them the
score atoms: the intercept, which is the score $1$ of the normalization, and
the atom $\tilde s$ of Eq.~\eqref{eq:1d-dict}, a multiple of the scale score $s_s$. With
the likelihood fit, $\psi=s$, so each score atom is a multiple of one
estimating function, $a=c\,\psi_m$. Since
$\mathbb E^{\rm tr}_{k,\hat\nu^{(-k)}}[\psi_m s^\top]$ is row $m$ of
$J_k-\Sigma_A^{-1}$ by Eq.~\eqref{eq:Jgen},
\[
\tilde a_k=c\,e_m^\top\Sigma_A^{-1}J_k^{-1}\psi=(1-\rho_{k,m})\,a ,
\]
where the second equality uses that $J_k$ is diagonal, the two score atoms
being orthogonal under the fitted null. In the unit-information
parametrization
\[
\rho_{k,m}=\frac{E^{\rm tr}_k}{E^{\rm tr}_k+\sigma_m^{-2}},
\]
with $E^{\rm tr}_k$ the expected number of training events
(Appendix~\ref{app:impl-vfit}) and $\sigma_m$ the uncertainty of the auxiliary measurement of nuisance $m$.
$\rho_{k,m}$ is the shrinkage of the constrained nuisance fit:
the fitted nuisance moves the fraction $\rho_{k,m}$ of the way from $\hat\nu_A$ to
the unconstrained estimate, and the witness's coefficient on the score atom is
$1-\rho_{k,m}$ times what it would be against an unconstrained null.

For the dictionary witness, $D_k$ in Eq.~\eqref{eq:Vfit} has the entry
$1-\rho_{k,m}$ for the score atom of nuisance $m$ and $1$ for the smooth atoms
and the bumps, which are not damped because they are orthogonal to $1$ and
$s_s$ under the fitted null by construction (Section~\ref{sec:1d-setup}).

For the network the residualization has no closed form: the parameter
gradients $a_k$ are not orthogonal to the scores, and the part of each that the
nuisance fit absorbs is not computed. Only the output bias, which moves the
witness along the constant, is damped by $1-\rho_{k,n}$; every other free
parameter keeps the factor $1$, although hidden units can also produce
near-constant shifts and the scale direction has no dedicated parameter.
Leaving a partly absorbed direction undamped over-counts the fitting variance,
so this choice is conservative.

\subsection{Other details of \texorpdfstring{$\hat V_{\rm fit}$}{V\_fit}}\label{app:impl-nn}
The networks are trained with Adam at learning rate $10^{-3}$, $\beta_1=0.9$,
$\beta_2=0.99$ and $\epsilon=10^{-7}$, as in Ref.~\cite{DAgnolo2021learning}, full
batch (one step per epoch), with the clipping applied after every step.
With $a_k(x)=\partial f(x;\theta)/\partial\theta$ restricted to the free
parameters, the curvature is the Gauss--Newton matrix of Eq.~\eqref{eq:wloss},
\begin{equation}\label{eq:GN}
H_k=\frac{K-1}{K}\int\hat n_k\,e^{\phi_k}\,a_ka_k^\top+\epsilon_H\,\overline{\mathrm{diag}H}\,I ,
\end{equation}
where $I$ is the identity and $\overline{\mathrm{diag}H}$ the mean diagonal
element of the first term, and the overlaps $\tilde\Omega_{kj}$ are built from
$a_k$ and $a_j$ on a common measure. Below we discuss
 the choice of free parameters, the ridge
$\epsilon_H$, and tapered witnesses.

\paragraph{Free parameters.} The free parameters are those strictly inside the
clipping bound, together with any parameters that are not clipped. A parameter
at the bound of a box constraint does not respond to a small change of the data
at an exact constrained optimum; training with a fixed number of epochs does
not reach one. In the null studies the rule can under-count the fitting variance
as more parameters reach the bound (Table~\ref{tab:sens-5d}: the variance of
$Z$ grows with the training length), and it is
adequate at the configurations we use.

\paragraph{Ridge.} Without regularization $H_k$ is nearly singular in some free
directions and $\hat V_{\rm fit}$ is unbiased on average but noisy, and
occasionally negative. The ridge $\epsilon_H=10^{-4}$ was chosen once, on null
simulations of the dilepton benchmark at an earlier configuration ($c=0.72$,
2800 epochs; 288 data sets), checked on held-out ones, and is used unchanged at
every configuration in both benchmarks; it is the only calibration constant
fixed in this way.

\paragraph{Tapering.} For a feature with a sparsely populated upper tail, the
taper of Section~\ref{sec:witness-nn} is
\begin{equation}\label{eq:taper}
\tau_k(x)=\min\Big\{1,\ \frac{E^{\rm tr}_k\,\big[1-F_k(x)\big]}{\kappa}\Big\},
\end{equation}
where $F_k$ is the distribution function of the fitted null and
$E^{\rm tr}_k$ its expected number of training events
(Appendix~\ref{app:impl-vfit}), so the numerator is the expected number of training
events beyond $x$. The witness $\tau_kf$ equals the network output while at
least $\kappa$ training events are expected beyond $x$ and fades in proportion
to that number further out. With a taper (used for the network of
Section~\ref{sec:1d}), the network responds to its
training events through the untapered gradient $a_k$, since it is trained
without the taper, while the statistic sees the tapered witness through
$\tau_ka_k$. The overlaps therefore pair one with the other:
$\Omega_{kj}$ in Eq.~\eqref{eq:Qstar} becomes
$\Omega_{kj}(\tau)=\int a_k\,(\tau_ja_j)^\top dQ^*$, computed on the common
measure $\hat Q$ as in Appendix~\ref{app:impl-vfit}, which reduces to
$\Omega_{kj}$ when $\tau\equiv1$. The dependence of $\tau_k$ on the nuisance
estimate is neglected; its contribution is the product of the witness and the
nuisance error, which is second order under the null hypothesis, where the
witness is itself of the size of the noise.
\section{Additional results}\label{app:extra}

This appendix collects results referred to in Sections~\ref{sec:1d}
and~\ref{sec:5d}: the null distribution of $Z$ for the dictionary witness, and
the full tuning grids behind Figs.~\ref{fig:1d-tuning} and~\ref{fig:5d-tuning}.

\begin{figure}[t]
\centering
\includegraphics[width=\textwidth]{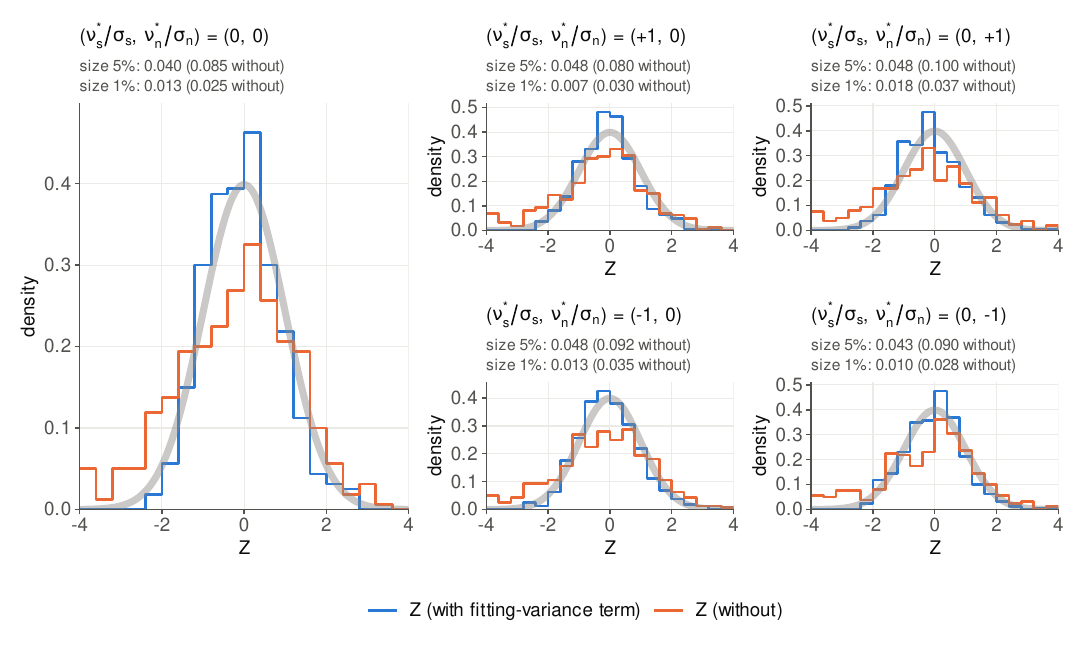}
\caption{As Fig.~\ref{fig:1d-null}, for the dictionary witness.}
\label{fig:1d-null-dict}
\end{figure}

\paragraph{Dictionary witness under the null hypothesis.}
Figure~\ref{fig:1d-null-dict} is the counterpart of Fig.~\ref{fig:1d-null} for
the dictionary witness, at the same five nuisance points with
$\sigma_A=0.05$. The picture is the same as for the neural network. With the
fitting-randomness term the rate at nominal $5\%$ is between $0.0400$ and
$0.0475$; without it the distribution is visibly too wide and the rate is
between $0.080$ and $0.100$, $1.6$ to $2$ times the nominal value.

\paragraph{Tuning grids.} Tables~\ref{tab:sens-1d} and~\ref{tab:sens-5d} give
the values plotted in Figs.~\ref{fig:1d-tuning} and~\ref{fig:5d-tuning}, and,
for the one-dimensional benchmark, the scan of the dictionary's penalty level
summarized in Section~\ref{sec:1d-tuning}. For each witness all
configurations are run on the same data sets, so differences between its rows
are not inflated by independent Monte Carlo noise.

\begin{table}[t]
\centering
\caption{One-dimensional benchmark, $\nu^*=0$, $\sigma_A=0.05$: false-positive
rate at nominal $5\%$ and $1\%$, and median $Z$ for NP1--NP3 with its bootstrap
standard error in parentheses, 400 data sets per entry. Network: clipping bound
$c$ and number of training epochs; dictionary: penalty level $\lambda$.
$^\dagger$: the configuration used in Section~\ref{sec:1d}.}
\label{tab:sens-1d}
\small
\setlength{\tabcolsep}{4pt}
\begin{tabular}{llccccc}
\toprule
 & setting & \multicolumn{2}{c}{null: rate at} & \multicolumn{3}{c}{median $Z$} \\
\cmidrule(lr){3-4}\cmidrule(lr){5-7}
witness & $(c,\text{epochs})$ / $\lambda$ & $5\%$ & $1\%$ & NP1 & NP2 & NP3 \\
\midrule
network & $(3,2\times10^4)$ & 0.035 & 0.000 & 2.22 (0.16) & 1.28 (0.07) & 3.38 (0.12) \\
network & $(3,3\times10^4)$ & 0.040 & 0.000 & 2.35 (0.16) & 1.26 (0.08) & 3.40 (0.11) \\
network & $(3,5\times10^4)$ & 0.038 & 0.000 & 2.42 (0.13) & 1.28 (0.09) & 3.39 (0.11) \\
network & $(4,2\times10^4)$ & 0.033 & 0.003 & 2.11 (0.17) & 1.24 (0.08) & 3.62 (0.12) \\
network & $(4,3\times10^4)$$^\dagger$ & 0.038 & 0.003 & 2.41 (0.15) & 1.19 (0.09) & 3.62 (0.11) \\
network & $(4,5\times10^4)$ & 0.045 & 0.003 & 2.51 (0.14) & 1.21 (0.10) & 3.61 (0.10) \\
network & $(6,2\times10^4)$ & 0.040 & 0.005 & 2.19 (0.17) & 1.15 (0.09) & 3.90 (0.12) \\
network & $(6,3\times10^4)$ & 0.040 & 0.008 & 2.40 (0.14) & 1.18 (0.08) & 3.88 (0.13) \\
network & $(6,5\times10^4)$ & 0.045 & 0.010 & 2.58 (0.16) & 1.13 (0.07) & 3.84 (0.12) \\
\midrule
dictionary & $1.1$ & 0.045 & 0.013 & 2.97 (0.11) & 1.09 (0.08) & 3.42 (0.07) \\
dictionary & $1.3$$^\dagger$ & 0.040 & 0.013 & 3.08 (0.09) & 1.16 (0.08) & 3.43 (0.07) \\
dictionary & $1.5$ & 0.038 & 0.010 & 3.09 (0.09) & 1.21 (0.08) & 3.38 (0.08) \\
\bottomrule
\end{tabular}

\end{table}

\begin{table}[t]
\centering
\caption{Dilepton benchmark, network witness: variance of $Z$ under the null
hypothesis and its rate at nominal $5\%$ and $1\%$ ($\nu^*=0$, 400 data sets), and median $Z$ for $Z'$ and EFT
(160 data sets) with its bootstrap standard error in parentheses, against the
clipping bound $c$ and the number of training epochs. $^\dagger$: the
configuration used in Section~\ref{sec:5d}.}
\label{tab:sens-5d}
\small
\setlength{\tabcolsep}{4pt}
\begin{tabular}{llccccc}
\toprule
 & & \multicolumn{3}{c}{null} & \multicolumn{2}{c}{median $Z$} \\
\cmidrule(lr){3-5}\cmidrule(lr){6-7}
$c$ & epochs & ${\rm Var}(Z)$ & rate at $5\%$ & rate at $1\%$ & $Z'$ & EFT \\
\midrule
$0.7$ & $2000$ & 0.83 & 0.0400 & 0.0075 & 2.68 (0.10) & 6.78 (0.12) \\
$0.7$ & $3000$$^\dagger$ & 0.93 & 0.0575 & 0.0125 & 2.80 (0.13) & 7.00 (0.13) \\
$0.7$ & $5000$ & 0.93 & 0.0575 & 0.0075 & 2.86 (0.16) & 7.00 (0.12) \\
$1.0$ & $2000$ & 0.90 & 0.0450 & 0.0075 & 2.80 (0.13) & 8.02 (0.18) \\
$1.0$ & $3000$ & 1.06 & 0.0625 & 0.0100 & 2.79 (0.13) & 8.52 (0.25) \\
$1.0$ & $5000$ & 1.11 & 0.0650 & 0.0100 & 2.85 (0.11) & 8.51 (0.24) \\
$1.3$ & $2000$ & 0.92 & 0.0500 & 0.0000 & 2.78 (0.11) & 8.69 (0.22) \\
$1.3$ & $3000$ & 1.08 & 0.0400 & 0.0025 & 2.91 (0.14) & 9.27 (0.32) \\
$1.3$ & $5000$ & 1.16 & 0.0600 & 0.0075 & 2.92 (0.17) & 9.45 (0.32) \\
\bottomrule
\end{tabular}

\end{table}

\bibliographystyle{unsrtnat}
\bibliography{refs_verified,refs}
\end{document}